\documentclass[11pt,a4paper]{article}
\usepackage{hyperref}
\usepackage{amssymb,amsmath,epsfig,color,amsthm}
\usepackage{slashbox}
\usepackage{tikz}
\usetikzlibrary{shapes,decorations.markings}

\newcommand{\dual}[1]{\bar{#1}}
\renewcommand{\vec}[1]{\mathbf{#1}}
\DeclareMathOperator{\tr}{\text{tr}}
\DeclareMathOperator*{\Det}{Det}
\newcommand{\id}{\text{id}}
\newcommand{\U}{\mathrm{U}}
\newcommand{\SU}{\mathrm{SU}}
\newcommand{\SO}{\mathrm{SO}}
\newcommand{\SP}{\mathrm{SP}}
\newcommand{\G}{\mathrm{G}}
\newcommand{\GL}{\mathrm{GL}}
\newcommand{\g}{\mathfrak{g}}
\newcommand{\HHS}{H_{\text{HS}}}
\newcommand{\smfrac}[2]{\genfrac{}{}{}{1}{#1}{#2}}
\newcommand{\gl}{\mathfrak{gl}}
\newcommand{\su}{\mathfrak{su}}

\newcommand{\Ed}{E}
\newcommand{\Pd}{P}
\newcommand{\wBr}{\mathrm{WB}_{L}}

\newcommand{\symm}{\mathfrak{S}}

\newcommand{\Real}{\mathbb{R}}
\newcommand{\Complex}{\mathbb{C}}
\newcommand{\Integer}{\mathbb{Z}}

\newcommand{\cC}{\mathcal{C}}
\newcommand{\cH}{\mathcal{H}}  
\newcommand{\cJ}{\mathcal{J}}
\newcommand{\cN}{\mathcal{N}}
\newcommand{\cP}{\mathcal{P}}
\newcommand{\cV}{\mathcal{V}}

\newcommand{\bE}{\mathbb{E}}
\newcommand{\bH}{\mathbb{H}}  
\newcommand{\bI}{\mathbb{I}}  
\newcommand{\bL}{\mathbb{L}}
\newcommand{\bP}{\mathbb{P}}
\newcommand{\bQ}{\mathbb{Q}}
\newcommand{\bS}{\mathbb{S}}
\newcommand{\bT}{\mathbb{T}}

\title{
\vspace{-2cm}
  Infinite Matrix Product States for long-range SU(N) spin models}

\author{Roberto Bondesan
 and Thomas Quella\\[3mm]
\small
  Institute of Theoretical Physics, University of Cologne\\
  \small Z\"ulpicher Stra\ss{}e 77, D-50937 Cologne, Germany
\\[3mm]
  {\em\small Emails: Roberto.Bondesan@uni-koeln.de, Thomas.Quella@uni-koeln.de}
}

\date{}

\begin{document}
\maketitle
\begin{abstract}
  We construct 1D and 2D long-range $\SU(N)$ spin models as parent
  Hamiltonians associated with infinite matrix product states. The
  latter are constructed from correlators of primary fields in the
  $\SU(N)_1$ WZW model. Since the resulting groundstates are of
  Gutzwiller-Jastrow type, our models can be regarded as lattice
  discretizations of fractional quantum Hall systems. We then focus on
  two specific types of 1D spin chains with spins located on the unit
  circle, a uniform and an alternating arrangement. For an equidistant
  distribution of identical spins we establish an explicit connection
  to the $\SU(N)$ Haldane-Shastry model, thereby proving that the model
  is critical and described by a $\SU(N)_1$ WZW model.
  In contrast, while turning out to be critical as well, the
  alternating model can only be treated numerically.
  Our numerical results rely on a
  reformulation of the original problem in terms of loop models.
\end{abstract}
\setcounter{tocdepth}{2}
\tableofcontents

\section{Introduction}

  Long-range spin models such as the Gaudin model \cite{Gaudin:1976}
  or the Haldane-Shastry model
  \cite{Haldane:1988PhRvL..60..635H,Shastry:1988PhRvL..60..639S} have
  attracted the attention of physicists and mathematicians for a long
  period of time. In its original formulation, the Haldane-Shastry
  model describes the dynamics of $\SU(2)$ spins on a circle with
  inverse distance square interactions. It received a lot of attention
  due to its exact solvability and due to the form of its groundstate
  which is closely related to a bosonic Laughlin wavefunction at
  filling fraction $\nu=1/2$. The Haldane-Shastry model can be viewed
  as realizing a 1D analogue of a chiral spin liquid, with spinon
  excitations satisfying a generalized Pauli exclusion principle and
  obeying fractional statistics
  \cite{Haldane:1991PhRvL..67..937H}. Many of the remarkable
  properties of the Haldane-Shastry model have their origin in the
  existence of an infinite-dimensional Yangian symmetry
  \cite{Haldane:1992PhRvL..69.2021H}. The latter also allowed to
  identify its thermodynamic limit as the $\SU(2)$ WZW conformal field
  theory at level $k=1$ \cite{Bernard:1993JPhA...26.5219B} (see also
  \cite{Bouwknegt:1994PhLB..338..448B}). The Haldane-Shastry model
  admits obvious generalizations to symmetry groups such as the
  unitary series $\SU(N)$ \cite{Haldane:1992PhRvL..69.2021H} or its
  supersymmetric  analog  $\SU(M|N)$
  \cite{Haldane:1994cond.mat..1001H,BasuMallick:2006NuPhB.757..280B}.

  For our current work, there are two aspects of the $\SU(N)$
  Haldane-Shastry model that will be particularly important. First of
  all, it provides an efficient discretization of the $\SU(N)$ WZW
  conformal field theory at level $k=1$, where the scaling laws are
  not affected by logarithmic corrections. Secondly, wavefunctions of
  the groundstate as well as the excited states exhibit an intimate
  relation to the physics of fractional quantum Hall (FQH) systems,
  also for general values of $N$ \cite{Kawakami:1992PhRvB..46.3191K}.
  There are of course also differences: While the constituents of FQH
  systems are particles which are moving on a 2D surface, the degrees
  of freedom in the spin model are pinned to fixed discrete locations
  on a circle.

  The study of fractional quantum Hall systems is frequently based on
  the following intriguing dichotomy: One single chiral 2D conformal
  field theory (CFT) describes the two complementary aspects of the
  physical sample -- its bulk and its boundary. It is known since the
  work of Moore and Read, for instance, that chiral CFT correlators
  give rise to realistic trial wave functions for the groundstates of
  gapped chiral 2D states of matter \cite{Moore:1991ks}. Remarkably,
  chiral correlators also encode the anyonic statistics of
  quasi-particle excitations above the groundstate. Among other
  insights, this led to the theoretical prediction of quasi-particles
  with non-abelian statistics for the FQH state at filling fraction
  $\nu=5/2$. At the same time, the chiral CFT describing the bulk can
  be used to model the properties of the 1D gapless theory describing
  its chiral edge \cite{Wen:1992IJMPB...6.1711W}. The intimate
  relation between bulk and boundary is also visible in entanglement
  spectra which can be calculated from the groundstate wave function
  \cite{Li:PhysRevLett.101.010504}.

  Recently, the question whether chiral topological states of matter
  can be engineered systematically received renewed interest. This is
  partly due to prospects of simulating strongly correlated systems in
  optical lattices. On the other hand, one also requires efficient
  ways of capturing the topological properties of strongly correlated
  systems from a numerical point of view. In 1D, all properties of
  gapped states are well captured by matrix product states
  \cite{Perez-Garcia:2007:MPS:2011832.2011833}. However, the situation
  is by far less obvious in 2D. While simple tensor network
  realizations for non-chiral topological states such as the Kitaev
  model \cite{Verstraete:2006PhRvL..96v0601V} or the Levin-Wen models
  \cite{Buerschaper:2009PhRvB..79h5119B} have been known for some
  time, chiral topological phases resisted all attempts to find such
  representatives. By now, there is considerable evidence that chiral
  topological phases cannot be described in terms of tensor network
  states with finite bond dimension, at least if one insists on a
  gapped parent Hamiltonian with local interactions
  \cite{Dubail:2013arXiv1307.7726D,Wahl:2013arXiv1308.0316W}.

  An interesting approach to the construction of chiral topological
  phases using ``infinite matrix product states'' has recently been
  suggested by Cirac and Sierra \cite{Cirac:2010PhRvB..81j4431C} and
  elaborated in more detail by Nielsen, Cirac and Sierra in
  \cite{Nielsen:2011py}. The basic idea is to define a spin model in
  terms of the data of an associated WZW model
  \cite{Witten:1983ar,Knizhnik:1984nr} for specified locations of the
  spins on the complex plane. More precisely, the Hamiltonian is
  designed as to annihilate a specific set of WZW correlation
  functions which, in turn, are used to define the groundstate of the
  spin model. This can be achieved by employing the existence of null
  fields in the WZW model (see Section~\ref{sc:Philosophy} for
  details). The resulting state can be interpreted as an infinite
  matrix product state since the fields can be regarded as operators
  on an infinite dimensional Hilbert space, with the correlator
  replacing the usual trace.

  When carried out for $\SU(2)$ spins on a circle, the previous program
  gives rise to a slight generalization of the Haldane-Shastry model
  \cite{Nielsen:2011py}. The match becomes perfect when the spins are
  distributed equidistantly. Starting from the $\SU(2)$ WZW model at
  level $k=1$, the authors of \cite{Nielsen:2011py} thus succeeded in
  defining a 1D spin model which gave rise to the same WZW model in
  the thermodynamic limit.\footnote{Parent Hamiltonians with similar
    features discretizing the $\SU(2)$ WZW model (even at arbitrary
    level $k$) have also been proposed in
    \cite{Greiter:2011jp,Thomale:2012PhRvB..85s5149T}, starting from a
    slightly different perspective.} By now, this program has also
  been put into effect for the groups $\SO(N)$ and $\U(1)$
  \cite{Tu:2013PhRvB..87d1103T,Tu:2014NJPh...16c3025T} leading to
  conceptually similar results. In the latter case it was also
  possible to interpolate 2D Laughlin states at filling fraction
  $\nu=1/q$ from the lattice to
  the  continuum.
  
  One of the notable features of the construction presented in
  \cite{Nielsen:2011py} is its remarkable flexibility. The spins can
  be placed at arbitrary positions in the complex plane, including
  regular arrangements such as various types of 1D or 2D
  lattices. Moreover, the transformation behavior of the spins can be
  chosen at will, even independently on each site. It is then a
  natural question to which extent the usual dichotomy of FQH states
  applies to this new type of construction. In particular, one would
  like to know whether the thermodynamic limit of a 2D setup describes
  a chiral topological state of matter and how its properties -- e.g.\
  its anyonic excitations, its edge theory and its groundstate
  entanglement spectrum -- relate to the data of the WZW model
  initially put in. Turning one's attention to 1D setups, one may --
  similarly -- expect a flow to a 2D CFT but a priori there is no
  reason why it should be connected to the original WZW model.  One
  may even speculate (and investigate) whether potential chiral
  topological phases resulting from 2D setups can be engineered
  systematically by stacking layers of 1D critical chains.  A simple
  general answer to our previous questions can not be expected. The
  exploration of individual examples is therefore the method of choice
  to gain a better idea about the value and the limitations of the
  general method of infinite matrix product states.

  In this paper, we apply the construction of \cite{Nielsen:2011py} to
  the case of the $\SU(N)$ WZW model at level $k=1$. The main
  motivation for this extension is the additional degree of freedom
  that comes with the extension from $\SU(2)$ to a higher rank unitary
  symmetry. In particular, for $N\geq3$ the fundamental field is
  distinct from the anti-fundamental one. With two types of mutually
  dual representations at our disposal, we can then realize
  a family of anti-ferromagnetic spin models on bipartite lattices or
  study the effects of frustration. Moreover, one may expect
  additional types of spin interactions due to the existence of higher
  rank invariant tensors for $\SU(N)$. Apart from these conceptual
  points the generalization from $N=2$ to arbitrary values of $N$ is
  interesting since it is likely to relate to well-known FQH trial
  states such as the Halperin state \cite{Halperin:1983} or the
  non-abelian spin singlet (NASS) state
  \cite{Ardonne:1999PhRvL..82.5096A}. In both cases there is a close
  relation to the $\SU(3)$ WZW model.

  Our paper is organized as follows. In Section~\ref{sc:Construction}
  we review the basic philosophy of \cite{Nielsen:2011py} and,
  focusing on the fundamental and anti-fundamental representation, we
  implement it for the $\SU(N)_1$ WZW model. Our construction gives
  rise to families of long-range $\SU(N)$ spin models labeled by the
  types of spins and their location in the complex plane. Generically,
  the Hamiltonian involves a mixture of two- and three-spin
  interactions. We then discuss particular choices of spin
  configurations and the resulting simplifications. Finally, the WZW
  correlation functions determining the groundstates of our spin
  models are evaluated using free field representations. Spin models
  based on a single representation (the ``uniform case'') are
  discussed in more detail in Section~\ref{sc:Uniform}. We first
  rewrite the general Hamiltonian in terms of permutation operators,
  thereby making the model amenable to an efficient numerical
  treatment. Afterwards we show that the three-spin interactions
  decouple from the local dynamics if the spins are located on the
  circle. In the case of an equidistant distribution we manage to
  recover the $\SU(N)$ Haldane-Shastry model. Apart from providing a
  complete analytic solution to the model, this observation also
  allows to identify the thermodynamic limit of the chain as the
  $\SU(N)$ WZW model at level $k=1$ (plus generalized chemical
  potentials incorporating the square and the cube of the total
  spin).
 
  The situation is very different for the mixed spin models which are
  discussed in Section~\ref{sc:Mixed}. Now, the Hamiltonian can be
  rewritten in terms of permutations and generators of a
  Temperley-Lieb algebra. Moreover, in contrast to the uniform case,
  we have not been able to come up with any special configuration of
  spins which allows to decouple the three-spin interactions. As a
  consequence, the model appears to be intractable using purely
  analytic methods. In order to obtain an efficient numerical
  implementation we use that the Temperley-Lieb generators and the
  permutations generate a diagram algebra known as the walled Brauer
  algebra. The latter is the basis for a loop model reformulation of
  the original eigenvalue problem which is described in
  Section~\ref{sc:Loops}. The final part of this section deals with
  the exact diagonalization of alternating chains with an equidistant
  distribution of spins on a circle. Pushing the analysis to chain
  lengths of up to $L=18$ sites for several values of $N$, we are able
  to predict that the chain becomes conformal in the thermodynamic
  limit.  In addition, we identify part of the conformal spectrum and
  rule out the possibility of an $\SU(N)$ WZW model as the critical
  theory. The concluding Section~\ref{sc:Conclusions} summarizes our
  findings and points out potential directions of future research.

\section{\label{sc:Construction}The construction of long-range
  $\mathbf{SU(N)}$ spin chains}

  In this section we define $\SU(N)$ spin models involving the
  fundamental and anti-fundamental representation at arbitrary
  positions on the complex plane. The construction of the Hamiltonian
  is based on the $\SU(N)_1$ WZW model whose relevant properties are
  reviewed in detail. The groundstates of all models can be evaluated
  explicitly and are related to wavefunctions of Gutzwiller-Jastrow
  type as they appear in the physics of fractional quantum Hall
  systems.

\subsection{\label{sc:Philosophy}The basic philosophy}

  Let us consider the WZW model associated with a Lie group $\G$
  \cite{Witten:1983ar,Knizhnik:1984nr}. It defines a 2D conformal
  field theory with an infinite dimensional current algebra symmetry
  which renders the model exactly solvable.  According to the
  philosophy of \cite{Cirac:2010PhRvB..81j4431C,Nielsen:2011py} there
  is a natural way of associating a quantum mechanical lattice model
  to any (chiral) correlator
\begin{align}
  \label{eq:WZWcorrelator}
  \psi(z_1,\ldots,z_L)
  \ =\ \bigl\langle\psi_1(z_1)\cdots\psi_L(z_L)\bigr\rangle
  \ \ ,
\end{align}
  of WZW primary fields $\psi_i(z_i)$. These fields are inserted at
  arbitrary but fixed positions $z_i$ on the complex plane and they
  may be of different type (hence the subscript $i$). Since the chiral
  WZW model has a global symmetry $\G$, the fields $\psi_i(z_i)$ should
  be thought of as vector valued. Each of them transforms in an
  irreducible representation $\cH_i$ of $\G$.

  The associated lattice model is obtained by interpreting the numbers
  $z_i\in\Complex$ as corresponding to the location of spin operators
  $\vec{S}_i$ representing the infinitesimal action of $\G$ on the
  irreducible representation $\cH_i$. This allows one to define a
  quantum spin model on the Hilbert space
\begin{align}
  \cH\ =\ \cH_1\otimes\cdots\otimes\cH_L\ \ .
\end{align}
  The connection to the WZW theory is established by
  demanding that the groundstate of the spin model may actually be
  expressed in terms of the WZW correlator as
\begin{align}
  \label{eq:gs}
  |\psi\rangle
  \ =\ \sum_{\{q_i\}}\psi_{q_1\ldots q_L}(z_1,\ldots,z_L)\,|q_1\cdots q_L\rangle
  \ \in\ \cH\ \ .
\end{align}
  Here, the vectors $|q_i\rangle$ form a basis of the Hilbert space
  $\cH_i$ and the correlator carries the dual quantum numbers $q_i$
  with respect to the action of the Lie group $\G$. Since the
  correlator is $\G$-invariant, the state $|\psi\rangle$ necessarily
  needs to be a singlet under $\G$.

  At this stage of the discussion we did not yet specify the
  Hamiltonian. The latter can be obtained by defining algebraic
  operators $\cP_i$ associated with the sites $z_i$, such that
\begin{align}
  \cP_i\,\psi(z_1,\ldots,z_L)\ =\ 0\ \ .
\end{align}
  The state $|\psi\rangle$ is then automatically a zero energy
  groundstate of the Hamiltonian
  \cite{Cirac:2010PhRvB..81j4431C,Nielsen:2011py}
\begin{align}
  \label{eq:GeneralH}
  H\ =\ \sum_i\cP_i^\dag\cP_i\ \ ,
\end{align}
  which is hermitean and positive semi-definite. The sum over sites
  $i$ in \eqref{eq:GeneralH} is motivated by the desire that the
  Hamiltonian reflects the symmetries of the arrangement of sites
  (e.g.\ translation symmetry). In view of the construction, it is
  natural to interpret $|\psi\rangle$ as an infinite matrix product
  state and $H$ as the associated parent Hamiltonian. For general
  choices of the representations $\cH_i$, the state $|\psi\rangle$
  needs not be the only groundstate though. The number of groundstates
  is given by the dimension of the space of conformal blocks
  to which the chiral correlation function \eqref{eq:WZWcorrelator}
  belongs.\footnote{To put it differently: The correlation function
    $\psi(z_1,\ldots,z_L)$ is not uniquely defined by
    Eq.~\eqref{eq:WZWcorrelator} but it rather has a certain degree of
    arbitrariness related to different choices of fusion channels.}

  In practice, the operators $\cP_i$ are constructed employing the
  existence of null fields $\chi_i(z_i)$ associated with the fields
  $\psi_i(z_i)$ \cite{Nielsen:2011py}. The null field $\chi_i(z_i)$
  can be thought of as a descendant of the field $\psi_i(z_i)$ which
  vanishes identically. It can thus be obtained by acting with
  currents on the field $\psi_i(z_i)$. Then, by definition of the null
  field and after the use of Ward identities (see
  \eqref{eq:WardIdentity} below), one obtains a relation of the form
\begin{align}
  0\ =\
  \bigl\langle\psi_1(z_1)\cdots\chi_i(z_i)\cdots\psi_L(z_L)\bigr\rangle
   \ =\ \cP_i\,\bigl\langle\psi_1(z_1)\cdots\psi_L(z_L)\bigr\rangle
   \ \ .
\end{align}
  The art of constructing the lattice model is thus reduced to the
  explicit realization of the null fields $\chi_i(z_i)$ and the
  derivation of the operators $\cP_i$. We refrain from presenting
  further details of the general construction (see
  \cite{Nielsen:2011py}) and focus on the particular case of the
  $\SU(N)_1$ WZW model from now on. We finally wish to stress that the
  projectors $\cP_i$, the Hamiltonian $H$ and the groundstate
  $|\psi\rangle$ all explicitly depend on the choice of positions
  $z_i$ and on the choice of representations $\cH_i$. This dependence
  will be suppressed in our notation since these quantities are
  thought of as being fixed once and for all.

\subsection[The $\SU(N)_1$ WZW model and its null
vectors]{\label{sc:NullVectors}The $\mathbf{SU(N)_1}$ WZW model and
  its null vectors}

  The basic structure of any WZW model on a Lie group $\G$ is a current
  algebra extending the corresponding Lie algebra $\g$. Denoting the
  chiral currents by $J^a(z)$, the resulting symmetry can be compactly
  expressed in terms of the operator product expansion (OPE)
\begin{align}
  J^a(z)\,J^b(w)
  \ =\ \frac{k\,\kappa^{ab}}{(z-w)^2}+\frac{i{f^{ab}}_c\,J^c(w)}{z-w}\ \ .
\end{align}
  Here, the matrix $\kappa^{ab}$ describes a suitably normalized
  invariant form and the structure constants ${f^{ab}}_c$ of $\g$ are
  real valued. The quantity $k$ is a non-negative integer known as the
  level. The WZW primary fields $\psi_i(w)$ are labeled by irreducible
  representations of $\G$. In terms of the currents, they are
  characterized by the OPE
\begin{align}
  \label{eq:WZWPrimaries}
  J^a(z)\,\psi_i(w)
  \ =\ \frac{S_i^a\psi_i(w)}{z-w}\ \ .
\end{align}
  In this formula, we think of $\psi_i(w)$ as being vector valued and
  $S_i^a$ refers to the corresponding representation matrices.

  We now specialize all our considerations to the case of $\SU(N)$ and
  level $k=1$. At this particular level, each of the WZW primary
  fields is labeled by one of the $N-1$ fundamental weights. While our
  analysis can in principle be extended to other cases, this paper
  will only be concerned with primary fields transforming in either
  the fundamental representation $\cV$ or the anti-fundamental
  representation $\dual{\cV}$.\footnote{In terms of Dynkin labels one
    has $\cV=(1,0,\ldots,0)$ and $\dual{\cV}=(0,\ldots,0,1)$.} In
  other words, our goal is to construct a long range quantum spin
  model on mixed Hilbert spaces of the form
\begin{align}
  \label{eq:GeneralHilbertSpace}
  \cH\ =\ \cV^{L-\ell}\otimes\dual{\cV}^{\ell}\ \ .
\end{align}
  In the language of the previous subsection, the set of sites
  $\bL=\{1,\ldots,L\}=\bS\cup\dual{\bS}$ decomposes into two subsets
  $\bS$ and $\dual{\bS}$ such that $\cH_i=\cV$ if $i\in\bS$ and
  $\cH_i=\dual{\cV}$ if $i\in\dual{\bS}$. In what follows, it will be
  convenient to distinguish the two types of sites by means of a
  parity map $d_\bullet:\bL\to\Integer_2$ which satisfies $d_i=1$ for
  $i\in\bS$ and $d_i=0$ for $i\in\dual{\bS}$. Two physically
  particularly interesting setups correspond to the {\em uniform case}
  with $\ell=0$ and to the {\em alternating case} where $L$ is even
  and $\ell=L/2$. In these two cases the Hilbert spaces are given by
\begin{align}
  \cH_{\text{uniform}}
  \ =\ \cV^{\otimes L}
  \qquad\text{ or }\qquad
  \cH_{\text{alternating}}
  \ =\ \bigl(\,\cV\otimes\dual{\cV}\,\bigr)^{\otimes L/2}\ \ .
\end{align}
  In the former case one has $\bS=\bL$ and $d_i=1$ while in the
  latter case we choose $\bS=\{1,3,\ldots,L-1\}$ and
  $\dual{\bS}=\{2,4,\ldots,L\}$ together with $d_i=i\text{ mod }2$
  (the map $d_i$ then determines the parity of the site). While our
  notation suggests a uniform or alternating arrangement along a 1D
  chain, we are in principle still free to choose an arbitrary
  arrangement of spins at this level, including various types of 2D
  setups (possibly even with random locations). We wish to emphasize
  that the alternating setup is the most natural one for the
  description of anti-ferromagnetic spin models on bipartite
  lattices. Indeed, in that case $L$ only needs to be even in order to
  admit a singlet in the spectrum while it needs to be a multiple of
  $N$ in the uniform case.

  The next step towards the construction of the Hamiltonian is the
  discussion of the desired groundstate correlation function. Let us
  denote by $\psi(z)$ and $\dual{\psi}(z)$ the vector valued primary
  fields associated with the representations $\cV$ and $\dual{\cV}$
  and the corresponding representation matrices by $\vec{T}$ and
  $\dual{\vec{T}}$. We wish to emphasize that $\dual{\psi}$ does not
  refer to an anti-chiral field but merely to the dual
  representation. Since we are only dealing with chiral CFTs there
  should be no chance of confusion. In terms of the general setup the
  restriction to two types of fields means
\begin{align}
  \psi_i(z)
  \ =\ \begin{cases}
          \ \psi(z)&,\ i\in\bS\\[2mm]
          \ \dual{\psi}(z)&,\ i\in\dual{\bS}
        \end{cases}
  \qquad\text{ and }\qquad
  \vec{S}_i
  \ =\ \begin{cases}
          \ \bI^{i-1}\otimes\vec{T}\otimes\bI^{L-i}&,\ i\in\bS\\[2mm]
          \ \bI^{i-1}\otimes\dual{\vec{T}}\otimes\bI^{L-i}&,\ i\in\dual{\bS}\ \ .
        \end{cases}
\end{align}
  The representation matrices $\vec{T}$ and $\dual{\vec{T}}$ are
  related by transposition as $\dual{T}^a=-(T^a)^T$. The groundstate
  \eqref{eq:gs} of the desired spin system will be determined by the
  correlation functions \eqref{eq:WZWcorrelator} which, for the
  uniform and the alternating model, become
\begin{equation}
  \label{eq:SUNcorrelators}
\begin{split}
  \psi_{\text{uniform}}(z_1,\ldots,z_L)
  &\ =\ \bigl\langle\psi(z_1)\cdots\psi(z_L)\bigr\rangle
  \quad\text{ and }\quad\\[2mm]
  \psi_{\text{alternating}}(z_1,\ldots,z_L)
  &\ =\
  \bigl\langle\psi(z_1)\dual{\psi}(z_2)\cdots\psi(z_{L-1})\dual{\psi}(z_L)\bigr\rangle
  \ \ .
\end{split}
\end{equation}
  Both correlators can be evaluated exactly using a free field
  representation (see Section \ref{sc:Correlators}). However, in this
  section we are merely interested in finding the operators $\cP_i$
  that annihilate these correlators.

  It turns out that both of the fields $\psi(w)$ and $\dual{\psi}(w)$
  have null descendants on the first energy level. They are obtained
  by acting on the fields with the current algebra modes
\begin{align}
  \label{eq:CurrentMode}
  J_{-1}^a\ =\ \oint_0\frac{dz}{2\pi i}\,z^{-1}\,J^a(z)
\end{align}
  and performing a suitable projection. In order to find this
  projection we first list all candidate fields on the first energy
  level. Since the current $J^a(z)$ is transforming in the adjoint
  representation $\cJ$ (which is selfdual,
  $\cJ=\dual{\cJ}$) the potential fields are those in the tensor
  products $\cJ\otimes\cV$ and $\cJ\otimes\dual{\cV}$,
  respectively. The decomposition of these tensor products can be
  established using Young tableaux techniques and it reads\footnote{In
    terms of Dynkin labels one has $\cJ=(1,0,\ldots,0,1)$,
    $\Lambda=(0,1,0,\ldots,1)$ and $\cN=(2,0,\ldots,0,1)$. The
  entries are swapped for the dual representations.}
\begin{align}
  \label{eq:TP}
  \cJ\otimes\cV
  \ =\ \cV\oplus\cN\oplus\Lambda
  \qquad\text{ and }\qquad
  \cJ\otimes\dual{\cV}
  \ =\ \dual{\cV}\oplus\dual{\cN}\oplus\dual{\Lambda}\ \ .
\end{align}
  As can be inferred from a comparison of conformal dimensions
  ($h_{\cN}=h_{\cV}+1$) or from an explicit construction
  (cf.~\cite{FrancescoCFT}), the relevant null fields $\chi(w)$ and
  $\dual{\chi}(w)$ are associated with the representations
  $\cN$ and $\dual{\cN}$, respectively.

  Let us now focus our attention onto a fixed single site $i$, with an
  insertion of the field $\psi_i(z)$ of type either $\psi(z)$ (for
  $i\in\bS$) or $\dual{\psi}(z)$ (for $i\in\dual{\bS}$). It remains to
  construct the projector $\cP_i$ onto the space of null states which
  can be either of the form $\cN\subset\cJ\otimes\cV$ or
  $\dual{\cN}\subset\cJ\otimes\dual{\cV}$. These projections can
  easily be realized using the action of the quadratic Casimir
  operator $C_i$ on the relevant tensor product and its known
  eigenvalues $C_\bullet$ on the irreducible representations in its
  decomposition. These eigenvalues are identical for dual
  representations. In both cases one hence obtains
\begin{align}
  \label{eq:NullProjector}
  \cP_i
  \ =\ \frac{(C_i-C_{\cV})(C_i-C_{\Lambda})}{(C_{\cN}-C_{\cV})(C_{\cN}-C_{\Lambda})}\ \ ,
\end{align}
  as is obvious from restricting $C_i$ to any of the irreducible
  components appearing in \eqref{eq:TP}. In order to rewrite this
  expression in terms of spin operators we introduce matrices
  $\vec{t}$ for the adjoint representation and write
  $C_i=(\vec{S}_i+\vec{t})^2$.\footnote{Scalar products of spin
    operators are defined as
    $\vec{t}_1\cdot\vec{t}_2=t_1^a\,\kappa_{ab}\,t_2^b$. The square
    $\vec{t}^2$ is an abbreviation for $\vec{t}\cdot\vec{t}$.}  The
  Casimir eigenvalues $C_{\cV}$, $C_{\cN}$ and $C_{\Lambda}$ needed to
  evaluate the projector \eqref{eq:NullProjector} are summarized in
  Table~\ref{tab:Representations}. After some elementary algebra one
  then ends up with
\begin{align}
  \label{eq:Projector}
  \cP_i
  \ =\ \tfrac{1}{2(N+1)}\Bigl[(\vec{S}_i\cdot\vec{t})^2
       +(N+1)\vec{S}_i\cdot\vec{t}
       +N\Bigr]\ \ .
\end{align}

\begin{table}
\begin{center}
\begin{tabular}{cccccc}
  Symbol & Arises in & Dynkin label & Name & Casimir eigenvalue & Interpretation
  \\\hline\hline&&&\\[-1em]
  $0$ & $\cV\otimes\dual{\cV}$ & $(0,\ldots,0)$ & Trivial & $0$ &\\[2mm]
  $\cV$ & $\cJ\otimes\cV$ & $(1,0,\ldots,0)$ & Fundamental & $\frac{1}{N}(N^2-1)$ & Physical site\\[2mm]
  $\cJ$ & $\cV\otimes\dual{\cV}$ & $(1,0,\ldots,0,1)$ & Adjoint & $2N$ & Current modes\\[2mm]
  $\cN$ & $\cJ\otimes\cV$ & $(2,0,\ldots,0,1)$ &&
  $\tfrac{1}{N}(N+1)(3N-1)$ & Null field \\[2mm]
  $\Lambda$ & $\cJ\otimes\cV$ & $(0,1,0,\ldots,0,1)$ &&
  $\tfrac{1}{N}(N-1)(3N+1)$ & \\\hline\\[-1em]
  $\Xi$ & $\cV\otimes\cV$ & $(2,0,\ldots,0)$ &&
  $\frac{2}{N}(N-1)(N+2)$ \\[2mm]
  $\Upsilon$ & $\cV\otimes\cV$ & $(0,1,0,\ldots,0)$ &&
  $\frac{2}{N}(N-2)(N+1)$\\\hline
\end{tabular}
\caption{\label{tab:Representations}Representations and their Casimir
  eigenvalues. The duals are obtained by swapping the entries of the
  tuple. They have the same Casimir eigenvalues.}
\end{center}
\end{table}

  The expression for the projector \eqref{eq:Projector} can be
  simplified by noting that the unit matrix $\bI$ together with the
  $N^2-1$ spin matrices $\vec{S}_i$ span the full space of $N\times N$
  matrices available on site $i$. As a consequence, bilinears in
  $\vec{S}_i$ can be reduced in degree using the identity
\begin{align}
  \label{eq:SProduct}
  S_i^aS_i^b
  \ =\ \tfrac{1}{2}\bigl[S_i^a,S_i^b\bigr]
       +\tfrac{1}{2}\bigl\{S_i^a,S_i^b\bigr\}
  \ =\ \tfrac{i}{2}{f^{ab}}_c\,S_i^c
       -\tfrac{1}{2}(-1)^{d_i}{d^{ab}}_c\,S_i^c
       +\tfrac{1}{N}\kappa^{ab}\,\bI\ \ .
\end{align}
  In order to distinguish between the two types of representation
  matrices $\vec{T}$ and $\dual{\vec{T}}$ which could enter here we
  used the parity map $d_\bullet:\bL\to\Integer_2$ which was
  introduced below \eqref{eq:GeneralHilbertSpace}. Equation
  \eqref{eq:SProduct} can be read as the defining relation of the
  completely symmetric rank-three tensor ${d^{ab}}_c$ which, moreover,
  is traceless. More details about the definition and the properties
  of the tensors $f$, $d$ and $\kappa$ can be found in
  Appendix~\ref{ap:SUN}. Using the product formula \eqref{eq:SProduct}
  and the explicit matrices ${(t^a)^b}_c=-i{f^{ab}}_c$
  for the adjoint representation it is now possible to derive an
  ``irreducible'' formula for the projector
  \eqref{eq:Projector}. Employing the formulas listed in
  Appendix~\ref{ap:SUN} it is straightforward even though slightly
  lengthy to verify that
\begin{align}
  {\bigl[\vec{t}\cdot\vec{S}_i\bigr]^a}_b
  \ =\ -i{f^a}_{bc}\,S_i^c
  \qquad\text{ and }\qquad
  {\bigl[(\vec{t}\cdot\vec{S}_i)^2\bigr]^a}_b
  \ =\ \tfrac{iN}{2}\,{f^a}_{bc}\,S_i^c
        -\tfrac{N}{2}(-1)^{d_i}\,{d^a}_{bc}\,S_i^c
        +2\delta_b^a\,\bI\ \ .
\end{align}
  Adding up all contributions with the correct coefficients, we find
\begin{align}
  \label{eq:ProjectorReduced}
  {\cP_i^a}_b
  \ =\ -\tfrac{i}{4}\tfrac{N+2}{N+1}\,{f^a}_{bc}\,S_i^c
       -(-1)^{d_i}\tfrac{N}{4(N+1)}\,{d^a}_{bc}\,S_i^c
       +\tfrac{N+2}{2(N+1)}\delta_b^a\,\bI\ \ .
\end{align}
  We note that $\cP_i$ is a hermitean operator-valued matrix which
  satisfies the projector property $\cP_i^2=\cP_i$.

\subsection{Derivation of the quantum spin Hamiltonians}

  The projectors in the previous subsection may be used to construct
  operators that annihilate correlation functions of the form
  \eqref{eq:WZWcorrelator}. The starting point is the chiral
  correlator
\begin{align}
  \label{eq:NullCorrelator}
  0\ =\
  \bigl\langle\psi_1(z_1)\cdots\chi_i(z_i)\cdots\psi_L(z_L)\bigr\rangle
  \ \ ,
\end{align}
  which is obtained from \eqref{eq:WZWcorrelator} by replacing the
  field $\psi_i(z_i)$ on site $z_i$ by its associated null field
  $\chi_i(z_i)$. The null field $\chi_i(z_i)$ can be identified with the
  fields in the subspace $\cN\subset\cJ\otimes\cV$ (or
  $\dual{\cN}\subset\cJ\otimes\dual{\cV}$), where the tensor product
  is spanned by fields of the form $J_{-1}^a\psi_i(z_i)$. Formally,
  this amounts to replacing the matrices $t^a$ by the operators
  $J_{-1}^a$. Using the projector \eqref{eq:ProjectorReduced}, the
  previous equation may then be rewritten as
\begin{align}
  0
   \ =\
   {\cP_i^a}_b\,\bigl\langle\psi_1(z_1)\cdots\bigl[J_{-1}^b\psi_i(z_i)\bigr]\cdots\psi_L(z_L)\bigr\rangle
\ \ .
\end{align}
  We wish to stress that, while implicit, the operator $\cP_i$ still
  contains the spin operator $\vec{S}_i$ acting on the field
  $\psi_i(z_i)$ on site $i$. 
  Next we employ the affine Ward identity
\begin{align}
  \label{eq:WardIdentity}
  \bigl\langle\psi_1(z_1)\cdots\bigl[J_{-1}^b\psi_i(z_i)\bigr]\cdots\psi_L(z_L)\bigr\rangle
  \ =\ \sum_{j(\neq i)}\frac{S_j^b}{z_i-z_j}\,
       \bigl\langle\psi_1(z_1)\cdots\psi_i(z_i)\cdots\psi_L(z_L)\bigr\rangle
\end{align}
  in order to move the action of $J_{-1}^b$ to the other fields in the
  chiral correlation function. Here and below the sum is not performed
  over the indices appearing in parentheses. It originates from
  combining equation \eqref{eq:CurrentMode} for the modes of the
  current with the definition \eqref{eq:WZWPrimaries} of primary
  fields.  As a result, we can re-interpret the trivial equation
  \eqref{eq:NullCorrelator} as the following algebraic condition on
  the original correlator:
\begin{align}
  \label{eq:NullOperator}
  \cP_i^a\bigl(\{z_l\}\bigr)\,
         \bigl\langle\psi_1(z_1)\cdots\psi_L(z_L)\bigr\rangle
   \ =\ 0
  \qquad\text{ with }\qquad
  \cP_i^a\bigl(\{z_l\}\bigr)
  \ =\ \sum_{j(\neq i)}\frac{{\cP_i^a}_bS_j^b}{z_i-z_j}\ \ .
\end{align}
  We note in passing that these operators somewhat resemble the Gaudin
  Hamiltonians \cite{Gaudin:1976}. It should be stressed, however,
  that the operators \eqref{eq:NullOperator} still carry an adjoint
  index instead of merely implementing an $\SU(N)$-invariant spin-spin
  coupling.

  In order to build an $\SU(N)$-invariant operator from
  $\cP_i^a\bigl(\{z_l\}\bigr)$ we need to form bilinears and contract
  the index $a$. However, depending on the choice of parameters $z_i$
  this may not result in a hermitean operator. This situation may be
  cured by conjugating one of the two operators before performing the
  contraction \cite{Nielsen:2011py}. This procedure results in a
  family of Hamiltonians
\begin{align}
  \label{eq:NullHamiltonian}
  H\big(\{z_l\}\bigr)
  \ =\ \sum_k\cP_{k,a}^\dag\big(\{z_l\}\bigr)\cP_k^a\big(\{z_l\}\bigr)
  \ =\ \sum_k\sum_{i,j(\neq k)}
       \frac{S_i^a\cP_{k,ab}S_j^b}{(\bar{z}_k-\bar{z}_i)(z_k-z_j)}\ \ ,
\end{align}
  parametrized by the fixed but arbitrary positions $z_i\in\Complex$
  of the spins. During the substitution we used the property
  that the matrices $\cP_i$ of Eq.~\eqref{eq:ProjectorReduced}
  are hermitean projectors. By construction, the resulting
  Hamiltonians  \eqref{eq:NullHamiltonian}
  are hermitean, $\SU(N)$-invariant and positive
  semi-definite. In addition, they annihilate the wave function
  $\bigl\langle\psi_1(z_1)\cdots\psi_L(z_L)\bigr\rangle$ (assuming
  that the latter is non-trivial). For some purposes, it will be
  convenient to replace the operators $\cP_i^a\bigl(\{z_l\}\bigr)$ by
  expressions with a slightly modified dependence on the coordinates
  $z_i$. The precise details and the motivation for this substitution
  will be explained in Section~\ref{sc:Modifications}.

  In the case under investigation (with level $k=1$), the fusion of
  all WZW primary fields is abelian
  and described by the group $\Integer_N$. The representations $\cV$
  and $\dual{\cV}$ correspond to the charges $1$ and $-1$ (modulo
  $N$), respectively.  The space of conformal blocks
  corresponding to the general mixed setup described in
  \eqref{eq:GeneralHilbertSpace} is thus one-dimensional if
  $L-2\ell\equiv0\text{ mod }N$ and trivial otherwise. For the uniform
  spin model on the Hilbert space $\cV^{\otimes L}$ we therefore expect
  a unique zero energy groundstate if $L$ is a multiple of
  $N$.\footnote{For different $L$ the Hamiltonian of course still
    exists but its groundstate(s) have neither zero energy nor are
    they given in terms of the chiral
    correlators~\eqref{eq:WZWcorrelator}.}
  In contrast, for the alternating case there is always a unique zero
  energy groundstate.

\subsection{\label{sc:Modifications}Modifications of the Hamiltonian}

  It was shown in \cite{Nielsen:2011py} in the case of $\SU(2)$ that
  the Hamiltonian \eqref{eq:NullHamiltonian} is closely related to a
  Haldane-Shastry Hamiltonian provided one replaces the operators
  $\cP_i^a\bigl(\{z_l\}\bigr)$ by new operators of the form
\begin{align}
  \label{eq:NullOperatorC}
  \cC_i^a\bigl(\{w_l\}\bigr)
  \ =\ \sum_{j(\neq i)}w_{ij}\,{\cP_i^a}_b\,S_j^b
\end{align}
  and chooses the special values $w_{ij}=(z_i+z_j)/(z_i-z_j)$
  for the parameters. We now briefly discuss in which sense such a
  substitution is also possible for $\SU(N)$ and of what form such
  modifications may generally be.

  The possible alterations we have an mind are based on the following
  observation. Every singlet wave function $|\psi\rangle$ satisfies
  the equation (for arbitrary but fixed $i$)
\begin{align}
  \label{eq:ModAux}
  0\ =\ \sum_{j(\neq i)}{\cP_i^a}_b\,S_j^b\,|\psi\rangle\ \ .
\end{align}
  Once more it is straightforward to verify this equation using the
  relations summarized in Appendix~\ref{ap:SUN}. Any singlet solution
  to $\cP_i^a\bigl(\{z_l\}\bigr)|\psi\rangle=0$ will thus also be a
  solution to $\cC_i^a\bigl(\{w_l\}\bigr)|\psi\rangle=0$ for any
\begin{align}
  \label{eq:Generalw}
  w_{ij}= \frac{f(z_i)}{z_i-z_j}+g(z_i)\ \ ,
\end{align}
  and an arbitrary choice of $f(z)$ and $g(z)$.\footnote{The functions
    $f$ and $g$ can, in principle, also be chosen differently for each
    value of $i$.} The particular setup discussed in
  \cite{Nielsen:2011py} is based on the choice $f(z_i)=2z_i$ and
  $g(z_i)=-1$.

  The Hamiltonians we shall consider in this article are all of the
  form
\begin{align}
  \label{eq:NullHamiltonianC}
  H\ =\ \sum_k\cC_{k,a}^\dag\big(\{z_l\}\bigr)\,\cC_k^a\big(\{z_l\}\bigr)
   \ =\ \sum_k\sum_{i,j(\neq k)}\bar{w}_{ki}w_{kj}\,
        S_i^a\,\cP_{k,ab}\,S_j^b\ \ ,
\end{align}
  where $\cC_k^a\big(\{z_l\}\bigr)$ is defined in
  \eqref{eq:NullOperatorC} with parameters $w_{ij}$ as defined in
  \eqref{eq:Generalw}. A priori, it is not clear whether
  the transition from $w_{ij}=1/(z_i-z_j)$ to a more general setup is
  modifying the basic physical properties in the thermodynamic
  limit. This may concern a potential criticality of a 1D system or the
  statistics of anyonic excitations in a potential gapped chiral
  2D topological phase. Let us, however, stress that any singlet
  groundstate of the modified Hamiltonian is still unique (as a zero
  energy {\em singlet}) since the procedure above can of course always
  be reversed.\footnote{We note in passing that an undesired effect of
    passing from $w_{ij}=1/(z_i-z_j)$ to $w_{ij}=(z_i+z_j)/(z_i-z_j)$
    becomes visible for $L=2$ and $z_2=-z_1$. In that case $w_{12}=0$
    and the Hamiltonian \eqref{eq:NullHamiltonianC} vanishes
    identically.}

\subsection{\label{sc:Simplifications}Simplification of the
  general Hamiltonian}

  Before discussing particular setups we would like to simplify the
  general Hamiltonian of the form
  \eqref{eq:NullHamiltonianC}. Plugging in the concrete expression
  \eqref{eq:ProjectorReduced} for $\cP_{ab}$ we find
\begin{align}
  \label{eq:HGeneral}
  H\ =\ \sum_k\sum_{i,j(\neq k)}
        \bar{w}_{ki}w_{kj}\Bigl\{
        -\tfrac{i}{4}\tfrac{N+2}{N+1}\,f_{abc}\,S_i^aS_j^bS_k^c
        -\tfrac{N(-1)^{d_k}}{4(N+1)}\,d_{abc}\,S_i^aS_j^bS_k^c
        +\tfrac{N+2}{2(N+1)}\,\vec{S}_i\cdot\vec{S}_j
        \Bigr\}\ \ .
\end{align}
  In order to simplify this expression further, we split the
  sum into contributions with $i=j$ and others with $i\neq
  j$. Whenever $i=j$ we can use the identities
  \eqref{eq:SProduct} and \eqref{eq:TensorIdentities} in order to
  reduce the degree employing the relations
\begin{align}
  \label{eq:iik}
  f_{abc}\,S_i^aS_i^bS_k^c
  \ =\ iN\,\vec{S}_i\cdot\vec{S}_k
  \qquad\text{ and }\qquad
  d_{abc}\,S^a_iS_i^bS_k^c
  \ =\ -(-1)^{d_i}\tfrac{N^2-4}{N}\,\vec{S}_i\cdot\vec{S}_k\ \ .
\end{align}
  After some simple algebra this results in\footnote{The summation
    range in the second sum is an abuse of notation. What is meant is
    that all indices are different.}
\begin{equation}
\begin{split}
  H&\ =\ \sum_{i\neq k}
         |w_{ki}|^2\Bigl\{
         \tfrac{N+2}{4(N+1)}\bigl[N+(-1)^{d_i+d_k}(N-2)\bigr]
         \vec{S}_i\cdot\vec{S}_k
         +\tfrac{(N+2)(N-1)}{2N}
         \Bigr\}\\[2mm]
   &\qquad+\sum_{i\neq j\neq k}
         \bar{w}_{ki}w_{kj}\Bigl\{
         -\tfrac{i}{4}\tfrac{N+2}{N+1}\,f_{abc}\,S_i^aS_j^bS_k^c
         -\tfrac{(-1)^{d_i}N}{4(N+1)}\,d_{abc}\,S_i^aS_j^bS_k^c
         +\tfrac{N+2}{2(N+1)}\,\vec{S}_i\cdot\vec{S}_j
         \Bigr\}\ \ .
\end{split}
\end{equation}
  Splitting the contributions according to the order of the
  interaction, this can be rewritten as
\begin{align}
  \label{eq:HamiltonianSpinAlt}
  H
  &\ =\ \tfrac{(N+2)(N-1)}{2N}\sum_{i\neq j}
        |w_{ij}|^2
        +\tfrac{N+2}{4(N+1)}\sum_{i\neq j}
        \Bigl\{\bigl[N+(-1)^{d_i+d_j}(N-2)\bigr]|w_{ji}|^2
        +2\sum_{k(\neq i,j)}\bar{w}_{ki}w_{kj}
        \Bigr\}\,\vec{S}_i\cdot\vec{S}_j\nonumber\\[2mm]
  &\qquad+\tfrac{1}{4(N+1)}\sum_{i\neq j\neq k}
        \bar{w}_{ki}w_{kj}\Bigl\{
        -i(N+2)\,f_{abc}\,S_i^aS_j^bS_k^c
        -N(-1)^{d_k}\,d_{abc}\,S_i^aS_j^bS_k^c
        \Bigr\}\ \ .
\end{align}
  Using the symmetry properties of the tensors $f_{abc}$ and $d_{abc}$
  we can achieve a final simplification. Restricting the summation
  to $i<j<k$ and adding the missing permutations by hand, we can
  express the Hamiltonian in terms of the two quantities
\begin{equation}
\begin{split}
  \label{eq:OmegaSA}
  \Omega_{ijk}^T
  &\ =\ (-1)^{d_i}\bigl[\bar{w}_{ij}w_{ik}+\bar{w}_{ik}w_{ij}\bigr]
        +(-1)^{d_j}\bigl[\bar{w}_{jk}w_{ji}+\bar{w}_{ji}w_{jk}\bigr]
        +(-1)^{d_k}\bigl[\bar{w}_{ki}w_{kj}+\bar{w}_{kj}w_{ki}\bigr]\\[2mm]
  &\ =\ 2\,\text{Re}\Bigl[(-1)^{d_i}\bar{w}_{ij}w_{ik}
        +(-1)^{d_j}\bar{w}_{jk}w_{ji}
        +(-1)^{d_k}\bar{w}_{ki}w_{kj}\Bigr]\ \ ,\\[2mm]
  \Omega_{ijk}^A
  &\ =\ -i\bigl(\bar{w}_{ij}w_{ik}-\bar{w}_{ik}w_{ij}-\bar{w}_{ji}w_{jk}
        -\bar{w}_{kj}w_{ki}+\bar{w}_{jk}w_{ji}+\bar{w}_{ki}w_{kj}\bigr)\\[2mm]
  &\ =\ 2\,\text{Im}(\bar{w}_{ij}w_{ik}
        +\bar{w}_{jk}w_{ji}
        +\bar{w}_{ki}w_{kj})\ \ .
\end{split}
\end{equation}
  The superscripts stand for ``twisted (symmetrized)'' and
  ``anti-symmetrized'', respectively. With these definitions one
  immediately finds
\begin{align}
  \label{eq:HamiltonianSpinAllOmegaGeneral}
  H
  &\ =\ \tfrac{(N+2)(N-1)}{2N}\sum_{i\neq j}
        |w_{ij}|^2
        +\tfrac{N+2}{4(N+1)}\sum_{i\neq j}
        \Bigl\{\bigl[N+(-1)^{d_i+d_j}(N-2)\bigr]|w_{ji}|^2
        +2\sum_{k(\neq i,j)}\bar{w}_{ki}w_{kj}
        \Bigr\}\,\vec{S}_i\cdot\vec{S}_j\nonumber\\[2mm]
  &\qquad+\tfrac{1}{4(N+1)}\sum_{i<j<k}
        \Bigl\{(N+2)\,\Omega_{kij}^A\,f_{abc}\,S_i^aS_j^bS_k^c
        -N\,\Omega_{kij}^T\,d_{abc}\,S_i^aS_j^bS_k^c
        \Bigr\}\ \ .
\end{align}
  The resulting Hamiltonian involves long-ranged two-spin and
  three-spin interactions coupling every site with every other
  site. As it stands it is still valid for general choices of $w_{kl}$
  and parameters $z_k$ (see \eqref{eq:Generalw}), including 2D spin
  systems. For this reason, there is hardly any hope for succeeding
  with an analytical treatment beyond writing down the exact
  groundstate (see Section~\ref{sc:Correlators}). In contrast, drastic
  simplifications can be expected in case the quantities
  $\Omega_{ijk}^T$ and $\Omega_{ijk}^A$ both vanish or are at least
  constant (i.e.\ independent of the indices $ijk$). This precisely
  occurs for specific types of 1D setups which will now be discussed
  in more detail. In that case, the three-spin couplings can be
  rewritten in terms of the {\em total} spin, i.e.\ they basically
  decouple from the local dynamics.

\subsection{\label{sc:Setups}Discussion of special setups}

  The Hamiltonians derived in the previous subsection make sense for
  arbitrary parameters $z_i\in\Complex$ on the complex plane. Mostly,
  we will however be interested in quite particular spin locations
  which lead to considerable simplifications of the
  Hamiltonians. After briefly discussing general aspects of the
  freedom of choice we will present a few concrete and physically
  relevant examples that will be used in subsequent sections.

\subsubsection{\label{eq:SetupGeneral}General aspects}

  As we have seen in Section \ref{sc:Simplifications}, the two main
  parameters governing the complexity of the Hamiltonians are the
  quantities $\Omega_{ijk}^T$ and $\Omega_{ijk}^A$ that have been
  defined in Eq.~\eqref{eq:OmegaSA}. A general Hamiltonian of the form
  \eqref{eq:HamiltonianSpinAllOmegaGeneral} will always involve
  three-spin interactions. These three-spin interactions can, however,
  be rewritten in terms of the {\em total spin} in case the two
  quantities $\Omega_{ijk}^T$ and $\Omega_{ijk}^A$ defined in
  \eqref{eq:OmegaSA} are constant (or even vanishing). This leads to
  drastic simplifications and allows to relate the Hamiltonians
  \eqref{eq:HamiltonianSpinAllOmegaGeneral} to more familiar quantum
  systems such as the Haldane-Shastry model for specific choices of
  parameters. We shall present some physically relevant examples in
  the subsequent sections.

  Before diving into concrete models we wish to summarize a few
  general properties of the two assignments we shall mainly be
  concerned with. These are
\begin{align}
  \label{eq:wpossibilities}
  a)\quad w_{ij}
  \ =\ \frac{1}{z_i-z_j}
  \qquad\text{ and }\qquad 
  b)\quad w_{ij}
  \ =\ \frac{z_i+z_j}{z_i-z_j}\ \ .
\end{align}
  In both cases one has the property $w_{ij}=-w_{ji}$. In addition,
  there are a number of non-trivial identities which, however, depend
  on the particular case under consideration. In particular, case a)
  leads to
\begin{align}
  \label{eq:DeltaZero}
  \Delta_{ijk}
  \ =\ w_{ij}w_{ik}+w_{jk}w_{ji}+w_{ki}w_{kj}
  \ =\ 0
  \qquad\text{(case a))}\ \ .
\end{align}
  Similarly, case b) has the immediate but important consequence
\begin{align}
  \label{eq:DeltaOne}
  \Delta_{ijk}
  \ =\ w_{ij}w_{ik}+w_{jk}w_{ji}+w_{ki}w_{kj}
  \ =\ 1
  \qquad\text{(case b))}\ \ .
\end{align}
  It may be shown that the parametrization b) of $w_{kl}$ in terms
  of the variables $z_i$ is the unique solution to this equation
  \cite{Haldane:1992PhRvL..69.2021H}. In case b) it is, moreover,
  possible to simplify squares in view of the relation
\begin{align}
  \label{eq:Identitywz}
  w_{kl}^2
  \ =\ 1+4\,\frac{z_kz_l}{z_{kl}^2}
  \qquad\text{ with }\qquad
  z_{kl}\ =\ z_k-z_l
  \qquad\text{(for case b)}\ \ .
\end{align}
  The importance of the quantity $\Delta_{ijk}$ and the relations
  \eqref{eq:DeltaZero} and \eqref{eq:DeltaOne} stems from the fact
  that $\Omega_{ijk}^T$ essentially reduces to $\Delta_{ijk}$ (up to
  the signs) for either purely real or purely imaginary values of
  $w_{ij}$ (such that $\bar{w}_{ij}$ can be replaced by $w_{ij}$ up to
  a sign). In both of these cases one, in addition, has
  $\Omega_{ijk}^A=0$, a relation which even holds for mixed spin
  models. We thus expect significant simplifications of the Hamiltonian
  \eqref{eq:HamiltonianSpinAllOmegaGeneral} for both of the cases
  \eqref{eq:wpossibilities} provided the parameters $w_{ij}$ satisfy
  these extra conditions.

  We conclude the general discussion with an analysis of conditions
  which enforce all $w_{kl}$ to be either real or imaginary for the
  two specific choices listed in \eqref{eq:wpossibilities}. We start
  with case a) where we find
\begin{align}
  \text{Re}(w_{kl})
  \ =\ \frac{\text{Re}(z_k-z_l)}{|z_k-z_l|^2}
  \qquad\text{ and }\qquad
  \text{Im}(w_{kl})
   \ =\ -\frac{\text{Im}(z_k-z_l)}{|z_k-z_l|^2}
  \qquad\text{(case a))}\ \ .
\end{align}
  As a consequence we expect simplifications in case a) if the spins
  are positioned along a horizontal or a vertical line. We note that
  the distance of these lines from the origin does not matter since it
  will cancel out when passing from $z_k$ to $w_{kl}$. In case b)
  of \eqref{eq:wpossibilities} one similarly obtains
\begin{align}
  \text{Re}(w_{kl})
  \ =\ \frac{|z_k|^2-|z_l|^2}{|z_k-z_l|^2}
  \qquad\text{ and }\qquad
  \text{Im}(w_{kl})
  \ =\ -2\,\frac{\text{Im}(z_k\bar{z_l})}{|z_k-z_l|^2}
  \qquad\text{(case b))}\ \ .
\end{align}
  The general solution to $\text{Re}(w_{kl})=0$ is thus
  $z_k=re^{i\theta_k}$, for arbitrary (real) values of
  $\theta_k$, i.e.\ the spins need to be located on a circle. Note
  that the resulting value of $w_{kl}$ does not depend on the choice
  of radius $r$. The general solution to $\text{Im}(w_{kl})=0$
  requires all the $z_k$ to have the same phase (up to $\pi$), i.e.\
  they should all be located on the same line through the
  origin. After having addressed potential simplifications in some
  detail we are now going to discuss particular setups in which they
  are realized.

\subsubsection{\label{sc:SetupCircle}Spins on the circle}

  As was found in \cite{Nielsen:2011py} for $\SU(2)$ and motivated more
  generally in Section~\ref{eq:SetupGeneral}, drastic simplifications
  occur if the spins are located on a circle
\begin{align}
  z_k\ =\ re^{i\theta_k}
  \qquad\text{ together with the choice }\qquad
  \label{eq:wmod}
  w_{kl}
  \ =\ \frac{z_k+z_l}{z_k-z_l}
  \ =\ -i\cot\tfrac{1}{2}(\theta_k-\theta_l)\ \ .
\end{align}
  In this case one obtains $\bar{w}_{kl}=-w_{kl}$ and hence
  $\Omega_{ijk}^A=0$. We note a possible relation to the trigonometric
  Haldane-Shastry model in view of the relation
\begin{align}
  \label{eq:HSzz}
  \frac{z_kz_l}{z_{kl}^2}
  \ =\ -\frac{1}{4\sin^2\frac{1}{2}(\theta_k-\theta_l)}\ \ .
\end{align}
  Also the equation $\Delta_{ijk}=1$ (see \eqref{eq:DeltaOne}) has a
  number of consequences. Using the antisymmetry $w_{kl}=-w_{lk}$, one
  for instance finds
\begin{align}
  \label{eq:Sumww}
  \sum_{k(\neq i,j)}w_{ki}w_{kj}
  \ =\ L-2+2w_{ij}^2-w_{ij}(\xi_i-\xi_j)
  \qquad\text{ with }\qquad
  \xi_i\ =\ \sum_{k(\neq i)}w_{ik}\ \ .
\end{align}
  Note that the convention for $\xi_i$ used here is different from the
  convention used for $c_i$ in \cite{Nielsen:2011py}.

\subsubsection{\label{sc:SetupEquidistant}Equidistant distribution of
  spins on the circle}

  For physical applications the most important choice of spin
  locations is the equidistant distribution on the circle. In the
  language of Section~\ref{sc:SetupCircle} this corresponds to
  $\theta_k=\tfrac{2\pi}{L}k$ such that one has
\begin{align}
  z_k\ =\ re^{\frac{2i\pi}{L}k}
  \qquad\text{ and }\qquad
  w_{kl}
  \ =\ -i\cot\frac{\pi}{L}(k-l)\ \ .
\end{align}
  One of the technical advantages of the equidistant distribution is
  the fact that certain summations can now be carried out
  explicitly. For instance one finds
\begin{align}
  \label{eq:xi_and_minusw}
  \xi_i
  \ =\ \sum_{k(\neq i)}w_{ik}\ =\ 0
  \qquad\text{ and (for even $L$)}\qquad
  \sum_{k(\neq i)}(-1)^k\,w_{ik}\ =\ 0\ \ .
\end{align}
  Other important sums which can be evaluated using these insights are
\begin{align}
  \label{eq:uniformww}
  \sum_{i\neq j}|w_{ij}|^2
  \ =\ \tfrac{1}{3}L(L-1)(L-2)
  \qquad\text{ and }\qquad
  \sum_{i\neq j\neq k}\bar{w}_{ki}w_{kj}
  \ =\ -\tfrac{1}{3}L(L-1)(L-2)\ \ .
\end{align}

\subsubsection{\label{sc:SetupLine}Spins on the real line}

  A case that was not studied in \cite{Nielsen:2011py} but which appears
  to be of similar interest is the case of real values $z_k=x_k\in\Real$
  together with the unmodified choice a) of $w_{kl}$ (see
\eqref{eq:wpossibilities}),
\begin{align}
  w_{kl}
  \ =\ \frac{1}{z_k-z_l}
  \ =\ \frac{1}{x_k-x_l}\ \ .
\end{align}
  This case leads to a significant simplification of the system since
  now $\Omega_{ijk}^T=\Omega_{ijk}^A=\Delta_{ijk}=0$ in the uniform
  case. For the Haldane-Shastry model, one may regard this setup as
  the classical limit of the chain on the circle, see
  \cite{Haldane:1994cond.mat..1001H} for a more detailed discussion of
  this point.

\subsubsection{\label{sc:SetupHyper}Hyperbolic case}

  Just for completeness we also briefly describe the hyperbolic case
  where the spin locations are chosen to be on the real line with
\begin{align}
  \label{eq:zhyper}
  z_k\ =\ re^{\omega_k}
  \qquad\text{ together with the choice }\qquad
  w_{kl}
  \ =\ \frac{z_k+z_l}{z_k-z_l}
  \ =\ \coth\tfrac{1}{2}(\omega_k-\omega_l)\ \ .
\end{align}
  In this equation, the $\omega_k$ are meant to be arbitrary real
  parameters. Formally, the assignment \eqref{eq:zhyper} corresponds
  to the choice $\theta_k=-i\omega_k$ in the discussion of
  Section~\ref{sc:SetupCircle}. Correspondingly, Eq.~\eqref{eq:HSzz}
  now gets replaced by
\begin{align}
  \frac{z_kz_l}{z_{kl}^2}
  \ =\ \frac{1}{4\sinh^2\frac{1}{2}(\omega_k-\omega_l)}\ \ .
\end{align}
  Even though of limited physical interest we decided to include this
  case since the associated Haldane-Shastry model exhibits a Yangian
  symmetry \cite{Haldane:1992PhRvL..69.2021H}. For the Yangian
  symmetry to be present one needs to work with a uniform setup and
  spin locations $\omega_k=\alpha k$ for some arbitrary constant
  $\alpha\in\Real$. In particular, this requires an infinite number
  of sites right from the very beginning.

\subsection{\label{sc:Correlators}Groundstate wavefunctions}

  We have seen in Section~\ref{sc:NullVectors} that the groundstates
  of the Hamiltonians \eqref{eq:HamiltonianSpinAllOmegaGeneral} are
  given in terms of WZW correlators \eqref{eq:SUNcorrelators}. We will
  now calculate these correlators using the vertex operator
  realization of $\SU(N)_1$. In the alternating case we shall also
  employ a free fermion construction. The resulting wave functions are
  always of Gutzwiller-Jastrow type.

\subsubsection{\label{eq:VertexConstruction}Vertex operator
  construction}

  The correlation functions \eqref{eq:SUNcorrelators} entering the
  groundstate of our physical system can be evaluated explicitly,
  thanks to the fact that the $\SU(N)_1$ WZW model is equivalent to a
  free field theory. Indeed, it simply corresponds to a system of
  $N-1$ free bosons which are compactified on the root lattice of
  $\su(N)$. Accordingly, the WZW currents and also the WZW primary
  fields can be expressed in terms of these free fields. The $N-1$
  Cartan operators just correspond to derivatives of the $N-1$ free
  bosons. On the other hand, vertex operators are required to
  represent root operators and primary fields. While the corresponding
  vertex operators can easily be identified on the basis of their
  conformal dimension, there is a certain subtlety regarding cocycle
  phase factors which are need to ensure the correct statistics of
  fields. Since in our approach relative phases have a drastic
  influence on the state  \eqref{eq:WZWcorrelator},  it is important to
  get these phases right.

  Let us start with defining a multi-component chiral bosonic field
  $\varphi^i(z)$ using the OPE
\begin{align}
  \varphi^i(z)\,\varphi^j(w)
  \ =\ -\delta^{ij}\,\ln(z-w)\ \ .
\end{align}
  The derivatives $H^i(z)=i\partial\varphi^i(z)$ generate a
  $\U(1)^{N-1}$ current algebra
\begin{align}
  H^i(z)\,H^j(w)
  \ =\ \frac{\delta^{ij}}{(z-w)^2}\ \ .
\end{align}
  The associated primary fields are vertex operators
\begin{align}
  V_\mu(z)\ =\ \,:e^{i\mu\cdot\varphi(z)}:
\end{align}
  which are labeled by $(N-1)$-tuples $\mu$ and which have the
  conformal dimension $h_\mu=\tfrac{1}{2}\mu^2$. For our purposes it
  will be useful to identify the tuple $\mu$ with weights of $\su(N)$.

  We recognize that the vertex operators $V_\alpha(z)$ associated with
  the roots $\alpha$ have conformal dimension $h=1$ due to
  $\alpha^2=2$. They may be used to extend the free boson chiral
  algebra to $\SU(N)_1$. The concrete expression for the root
  generators is
\begin{align}
  E^\alpha(z)\ =\ c_\alpha V_\alpha(z)
\end{align}
  where $c_\alpha$ is a $\Integer_2$-valued cocycle ensuring the
  correct statistics of the currents (see e.g.\
  \cite{FrancescoCFT}). In our context, more important than the
  currents are the WZW primary fields associated with the
  fundamental and the anti-fundamental representation. The latter are
  known to have conformal dimension
  $h_\cV=h_{\dual{\cV}}=\tfrac{N-1}{2N}$. They are realized in terms
  of vertex operators $V_\mu(z)$ where $\mu$ is any weight of the
  corresponding representations. Indeed, the length of the
  corresponding weights is given by $\mu^2=\tfrac{N-1}{N}$, in
  accordance with our previous claim about the conformal
  dimension. The fundamental WZW primaries admit a representation as
\begin{align}
  \label{eq:FundFreeFields}
  \psi_q(z)
  \ =\ c_{\mu(q)}\,V_{\mu(q)}(z)
  \qquad\text{ and }\qquad
  \dual{\psi}_q(z)
  \ =\ c_{\bar{\mu}(q)}\,V_{\bar{\mu}(q)}(z)
\end{align}
  with another cocycle $c_\mu$ (see
  \cite{Chu:1994un,FrancescoCFT}). The cocycle depends on the indices
  $q$ through their respective weight $\mu(q)$ or $\bar{\mu}(q)$ in
  the fundamental or anti-fundamental representation. The latter may
  be written as
\begin{align}
  \mu(q)
  \ =\ \omega_1 - \sum_{r=1}^{q-1} \alpha_r
  \qquad\text{ and }\qquad
  \bar{\mu}(q)
  \ =\ \omega_{N-1} - \sum_{r=1}^{q-1} \alpha_{N-r}\ \ .
\end{align}
  In this formula, $\alpha_i$ denote the simple roots of $\su(N)$ while
  $\omega_1$ and $\omega_{N-1}$ refer to the highest weights of the
  fundamental and anti-fundamental representations $\cV$ and
  $\dual{\cV}$.

  The primary fields $\psi_q(z)$ and $\dual{\psi}_q(z)$ define
  correlation functions of the form \eqref{eq:SUNcorrelators} and
  thereby the desired quantum state
\begin{align}
  \label{eq:DesiredState}
  |\psi\rangle
  \ =\ \sum_{\{q_i\}}\psi_{q_1\cdots q_L}(z_1,\dots,z_L)
       \,|q_1\cdots q_L\rangle\ \ .
\end{align}
  Up to a coordinate independent sign stemming from the cocycles
  $c_\mu$, all relevant correlation functions can easily be calculated
  using the free field expression
\begin{align}
  \bigl\langle V_{\mu_1}(z_1)\cdots V_{\mu_L}(z_L)\bigr\rangle
  \ =\ \delta_{\mu,0}\,\prod_{i<j}(z_i-z_j)^{\mu_i\cdot\mu_j}\ \ .
\end{align}
  The Kronecker delta $\delta_{\mu,0}$ with $\mu=\sum_i\mu_i$ results
  from charge conservation. In the uniform case, it forces all
  correlation functions to vanish except if $L$ is a multiple of
  $N$. In the alternating case, it is sufficient for $L$ to be
  even. We shall now present two alternative ways for the explicit
  construction of the states \eqref{eq:DesiredState}.

\subsubsection{\label{sc:SignFactors}Determination of the sign
  factors}

  Let us focus on the uniform case first where all fields transform in
  the fundamental representation. The relevant correlation function
  then reads
\begin{align}
  \label{eq:CorrUpToSign}
  \psi_{q_1\cdots q_L}(z_1,\dots,z_L)
  \ =\  \delta_{\mu(q),0}\,e^{if(\{ q_i \})}
        \prod_{i<j}(z_i-z_j)^{\mu(q_i)\cdot \mu(q_j)}\ \ ,
\end{align}
  and the non-trivial task consists in determining the sign factor
  $e^{if(\{q_i\})}$. There are at least three distinct ways of
  accomplishing this. First of all, the sign can be determined through
  a detailed analysis of the cocycles entering the definition
  \eqref{eq:FundFreeFields} of the WZW primaries. Alternatively, the
  relative signs are fixed by the invariance of the state
  \eqref{eq:DesiredState} under the global action of $\SU(N)$. Here we
  shall follow an even simpler route which has been suggested in
  \cite{Nielsen:2012zb}. It employs the fact that singlet
  wavefunctions of the type \eqref{eq:CorrUpToSign} previously arose
  in the context of the $\SU(N)$ Haldane-Shastry model
  \cite{Kawakami:1992PhRvB..46.3191K}. There is just one slight
  difference to our setup: We are interested in general locations of
  the spins while the coordinates $z_k$ are distributed uniformly on
  the circle for the Haldane-Shastry model, $z_k=e^{2\pi i
    k/L}$. Since the sign factors in \eqref{eq:CorrUpToSign} do not
  depend on these coordinates  they may nevertheless be obtained by
  means of a simple comparison.

  The groundstate of the $\SU(N)$ Haldane-Shastry model is described in
  terms of the wave function \cite{Kawakami:1992PhRvB..46.3191K}
\begin{align}
  \label{eq:psiG}
  \psi_G\bigl(n_i^{(a)}\bigr)
  \ =\ e^{-i\pi\sum_{i,a}n_{i}^{(a)}}
       \prod_{a,i<j}D\Bigl(n_i^{(a)}-n_j^{(a)}\Bigr)^2
       \prod_{a<b,i,j}D\Bigl(n_i^{(a)}-n_j^{(b)}\Bigr)
\end{align}
  with $D(x-y)=\sin\bigl(\pi(x-y)/L\bigr)$. Here $a=2,3,\dots,N$ and
  $n_i^{(a)}$ is the position of the $i$-th site with spin $a$. The
  state is obtained from a Gutzwiller projection and hence a singlet
  by construction. Assuming neutrality of the configurations and
  noting that $\mu(q)^2=(N-1)/N$ for every $q=1,\dots,N$, one gets
\begin{equation}
\begin{split}
  &\prod_{n<m}
  (z_n-z_m)^{\mu(q_n)\cdot \mu(q_m)}=
  \prod_{n<m}
  \Bigl[2 i e^{i\pi (n+m)/L} D(n-m)\Bigr]^{\mu(q_n)\cdot \mu(q_m)}\\[2mm]
  &\ =\ 
  e^{\frac{1}{2}(\sum_{n,m}-\sum_{n,m}\delta_{mn})
    (\log(2i) + \frac{i\pi}{L}(n+m))\mu(q_n)\cdot \mu(q_m)
  }
  \prod_{n<m}
  D(n-m)^{\mu(q_n)\cdot \mu(q_m)}\\[2mm]
  &\ =\   
  e^{-\frac{N-1}{2N}
    (\log(2i) L + i\pi(L+1))
  }
  \prod_{n<m}
  D(n-m)^{\mu(q_n)\cdot \mu(q_m)}
  \equiv C_L    \prod_{n<m}
  D(n-m)^{\mu(q_n)\cdot \mu(q_m)}\, .
\end{split}
\end{equation}
  Relating this expression to the Gutzwiller wavefunction
  \eqref{eq:psiG} is a simple exercise. Note first the
  following fact:
\begin{align}
  \bigl(\mu(p)-\omega_1\bigr)\cdot\bigl(\mu(q)-\omega_1\bigr)
  \ =\ \sum_{r=1}^{p-1}\sum_{s=1}^{q-1}A_{rs}
  \ =\ \begin{cases}
         0 & \text{if } q=1 \text{ or } p=1\\[2mm]
         2 & \text{if } q=p\neq 1\\[2mm]
         1 & \text{otherwise}\ \ ,
       \end{cases}
\end{align}
  where $A_{rs}$ is the Cartan matrix of $\su(N)$. This allows us to
  rewrite $\psi_G$ as
\begin{align}
  \psi_G\bigl(n_i^{(a)}\bigr)
  \ =\ \delta_{\mu(q),0}\,
  e^{-i\pi\sum_{i,a}n_{i}^{(a)}}
  \prod_{n<m}
  \Bigl[(-1)^{\theta(q_n-q_m)} 
    D(n-m)\Bigr]^{(\mu(q_n)-\omega_1)\cdot(\mu(q_m)-\omega_1)}\ \ .
\end{align}
  The step function satisfies $\theta(p-q)=1$ if $p>q$ and $0$
  otherwise. It arises from the condition $a<b$ in
  Eq.~\eqref{eq:psiG}, meaning that if $q_n>q_m$, that term has an
  additional minus sign. Note that if $N=2$ the exponent is even, so
  that this extra sign is not  present.  Still assuming charge neutrality
  we now further note that
\begin{equation}
\begin{split}
  & \prod_{n<m}D(n-m)^{-\omega_1\cdot (\mu(q_n)+\mu(q_m))}
   \ =\ \prod_{n>m} (-1)^{-\omega_1\cdot \mu(q_n)}
        \prod_{n\neq m} D(n-m)^{-\omega_1\cdot \mu(q_n)}\\[2mm]
  &\ =\ \prod_{n} (-1)^{-\omega_1\cdot \mu(q_n) (n-1)}
        \prod_{n} \Bigl[ (-1)^{L-n} 2^{1-L} \, L
        \Bigr]^{-\omega_1\cdot \mu(q_n)}
   \ =\ 1\ \ .
\end{split}
\end{equation}
  As a consequence, the wavefunction simplifies to
\begin{equation}
\begin{split}
  \psi_G\bigl(n_i^{(a)}\bigr)
  &\ =\ \tilde{C}_L\,\delta_{\vec{q},0}\,
        e^{-i\pi\sum_{i,a}n_{i}^{(a)}}\prod_{n<m}
        (-1)^{\theta(q_n-q_m) (\mu(q_n)-\omega_1)\cdot (\mu(q_m)-\omega_1)}\\[2mm]
  &\qquad\times \prod_{n<m}D(n-m)^{\mu(q_n)\cdot \mu(q_m)}\ \ ,
\end{split}
\end{equation}
  where we introduced the constant
  $\tilde{C}_L=\prod_{n<m}D(n-m)^{\omega_1^2}$ and
  $\omega_1^2=(N-1)/N$. The sign factor may now be fixed by demanding
  that the previous expression equals the chiral correlator
  \eqref{eq:CorrUpToSign} when $z_n=e^{2\pi i n/L}$. We then find
\begin{equation}
\begin{split}
  e^{i f(\{q_i \})}
  &\ =\ e^{-i\pi\sum_{i,a}n_{i}^{(a)}}
  \prod_{n<m}
  (-1)^{\theta(q_n-q_m) (\mu(q_n)-\omega_1)\cdot
    (\mu(q_m)-\omega_1)}\\[2mm]
  &\ =\ \prod_n e^{-i\frac{\pi}{2} n (\mu(q_n)-\omega_1)^2}
  \prod_{n<m}
  (-1)^{\theta(q_n-q_m) (\mu(q_n)-\omega_1)\cdot
    (\mu(q_m)-\omega_1)}
  \ \ .
\end{split}
\end{equation}
  For the alternating setup involving correlation functions of both
  the fundamental and the anti-fundamental field the previous trick is
  not applicable and one would need to understand the cocycle
  properties in more detail. We shall employ a shortcut in
  that case, employing free fermions instead of free bosons.

\subsubsection{Free fermion construction}

  In the case of the alternating spin model there is an alternative
  perspective on the derivation of the groundstate wavefunction which
  we find worth mentioning. Namely, the fundamental representation
  $\cV$ of $\SU(N)$ can be interpreted as an $N$-dimensional representation of
  $\U(N)$ on which its $\U(1)$ subgroup acts trivially. This has to be
  distinguished from the fundamental representation $\cV_Q$ of $\U(N)$
  which carries a non-trivial $\U(1)$ charge $Q$. Similar arguments
  apply to the anti-fundamental representation $\dual{\cV}_Q$ which
  carries a $\U(1)$ charge $-Q$. Since the $\U(1)$ charges simply add up
  in tensor product, one has the identity
  $\cV\otimes\dual{\cV}=\cV_Q\otimes\dual{\cV}_Q$ where both sides can
  now be regarded as representations of $\SU(N)$.\footnote{Note that we
  need the alternation in order to eventually reach a representation
  with vanishing $\U(1)$ charge, thereby allowing us to descend from
  the group $\U(N)$ to its quotient $\SU(N)$.}

  The previous arguments can be lifted to the level of WZW
  theories. The great advantage of this re-interpretation is that the
  $\U(N)_1$ WZW model admits a representation in terms of $N$ complex
  fermions with non-trivial OPE\footnote{We stress once more that the
    bar is used to denote the dual representation. All fields
    considered here are holomorphic.}
\begin{align}
  \Psi_p(z)\,\bar{\Psi}^q(w)
  \ =\ \frac{\delta_p^q}{z-w}
  \qquad(\text{with }p,q=1,\ldots,N)\ \ .
\end{align}
  The currents ${J_p}^q(z)=:\Psi_p\bar{\Psi}^q:(z)$ are simply
  bilinears in these fermions. More importantly, it is easy to verify
  that the fields $\Psi_p(z)$ and $\bar{\Psi}^q(z)$  are WZW primary
  fields with $h=1/2$ and that they correspond to the fundamental and
  anti-fundamental representation of $\U(N)$, respectively.

  From the perspective of the $\SU(N)_1$ WZW model one can reconstruct
  the $\U(N)_1$ WZW model by extending it by a free field $\varphi(z)$
  generating the extra $\U(1)$. The associated $\U(1)$ charges $\pm Q$
  are carried by vertex operators $V_\pm(z)$ of this bosonic field. In
  this language one can then realize the fundamental $\U(N)$ fields
  $\Psi_p(z)$ and $\dual{\Psi}^q(z)$ in terms of the fundamental
  $\SU(N)$ fields $\psi_p(z)$ and $\dual{\psi}^q(z)$ as
\begin{align}
  \label{eq:USUIdentification}
  \Psi_q(z)\ =\ V_+\psi_q(z)
  \qquad\text{ and }\qquad
  \bar{\Psi}^q(z)\ =\ V_-\bar{\psi}^q(z)\ \ .
\end{align}
  The primary fields $\psi_q$ and $\bar{\psi}_q$ of the $\SU(N)_1$
  WZW model have conformal dimension $h=\frac{N-1}{2N}$. In order to
  make up for the desired $h=1/2$, the difference needs to be carried
  by the vertex operator. This forces the latter to have the form
\begin{align}
  V_\pm(z)
  \ =\ :e^{\pm i\varphi/\sqrt{N}}:\ \ ,
\end{align}
  with conformal dimension $h_\pm=1/2N$.

  With the identification \eqref{eq:USUIdentification} one can now
  easily determine the desired correlation functions of fundamental
  and anti-fundamental fields. They are given by
\begin{equation}
  \psi_{q_1\ldots q_L}(z_1,\ldots,z_L)
  \ =\
  \bigl\langle\psi_{q_1}(z_1)\bar{\psi}_{q_2}(z_2)\cdots\bigr\rangle
  \ =\ \frac{\bigl\langle\Psi_{q_1}(z_1)\bar{\Psi}_{q_2}(z_2)\cdots\bigr\rangle}
  {\bigl\langle V_+(z_1)V_-(z_2)\cdots\bigr\rangle}\ \ .
\end{equation}
  The two correlation functions entering this expression can be
  calculated using Wick's Theorem for free fields. One obtains
\begin{align}
  \bigl\langle\Psi_{q_1}(z_1)\bar{\Psi}^{q_2}(z_2)\cdots
  \Psi_{q_{L-1}}(z_{L-1})\bar{\Psi}^{q_L}(z_L)\bigr\rangle
  &\ =\ \Det_{1\le i,j\le L/2}
        \left(
          \frac{\delta_{q_{2i-1}}^{q_{2j}}}{z_{2i-1}-z_{2j}} 
        \right)\\[2mm]
  \bigl\langle V_+(z_1)V_-(z_2)\cdots V_+(z_{L-1})V_-(z_L)\bigr\rangle
  &\ =\ \prod_{1\le i<j\le L} (z_i-z_j)^{(-1)^{i+j}/N}\ \ .
\end{align}
  The advantage of this representation of the correlation function is
  the absence of any cocycles which obscure the correct sign factors.

\section{\label{sc:Uniform}Discussion of the uniform spin models}

  This section will be used to illuminate the structure of the
  Hamiltonians \eqref{eq:HamiltonianSpinAllOmegaGeneral} in the
  uniform case. For 1D models with spins located on a circle we will
  recover a slight modification of the $\SU(N)$ Haldane-Shastry model.
  For the equidistant case this
  allows to come up with a complete analytic solution for the
  spectrum. As a byproduct we find that the thermodynamic limit of the
  spin chain is described by an $\SU(N)_1$ WZW model.

\subsection{Simplification of the Hamiltonian}

  The uniform spin model is defined in terms of the partition
  $\bS=\bL$ and $\dual{\bS}=\emptyset$. The degree map will be chosen
  such that $(-1)^{d_i}=-1$. The general Hamiltonian
  \eqref{eq:HamiltonianSpinAllOmegaGeneral} simplifies accordingly and
  becomes
\begin{equation}
  \label{eq:HamiltonianSpinAllOmega}
\begin{split}
  H
  &\ =\ \tfrac{(N+2)(N-1)}{2N}\sum_{i\neq j}
        |w_{ij}|^2
        +\tfrac{N+2}{2(N+1)}\sum_{i\neq j}
        \Bigl\{(N-1)|w_{ji}|^2
        +\sum_{k(\neq i,j)}\bar{w}_{ki}w_{kj}
        \Bigr\}\vec{S}_i\cdot\vec{S}_j\\[2mm]
  &\qquad+\tfrac{1}{4(N+1)}\sum_{i<j<k}
        \Bigl\{(N+2)f_{abc}\,\Omega_{kij}^A
        -Nd_{abc}\,\Omega_{kij}^T
        \Bigr\}S_i^aS_j^bS_k^c\ \ .
\end{split}
\end{equation}
  Apart from the simplified expression for $\Omega_{ijk}^T$ there is
  otherwise nothing we can achieve on this level of
  generality. Further simplifications, however, can be realized if we
  restrict our attention to special choices of the positions $z_i$ and
  of the associated parameters $w_{ij}$.

  The form \eqref{eq:HamiltonianSpinAllOmega} of the Hamiltonian is
  not particularly suitable for a numerical treatment, in particular
  for larger values of $N$, since it involves rather complicated sums
  over the spin indices. On the other hand, we know that all
  contributions correspond to $\SU(N)$-invariant operators on the
  tensor products $\cV\otimes\cV$ and $\cV\otimes\cV\otimes\cV$ of two
  and three physical sites, respectively. Fortunately, these operators
  are exhausted by the identity operator $\bI$, the two-site
  permutations $\bP_{ij}$ and the cyclic permutations $\bT_{ijk}$. In
  terms of these operators the numerical implementation becomes much
  more efficient. Eventually, the complexity of the diagonalization
  problem even becomes independent of the value of $N$, see
  Section~\ref{sc:Loops} for a more detailed discussion of this
  point. The labor to find explicit expressions for the individual
  terms entering \eqref{eq:HamiltonianSpinAllOmega} is the only prize
  we have to pay.

  For the transpositions the story is not too difficult, given the
  known Casimir eigenvalues in the decomposition of
  $\cV\otimes\cV=\Xi\oplus\Upsilon$, see Table~\ref{tab:Representations}.  
  Indeed, one may easily verify the relation
\begin{align}
  \bP_{ij}
  \ =\ \vec{S}_i\cdot\vec{S}_j+\tfrac{1}{N}\ \ .
\end{align}
  These operators provide a unitary representation of the permutation
  group, i.e.\ they satisfy $\bP_{ij}^\dag=\bP_{ij}^{-1}$ as well as
\begin{align}
  \label{eq:PermutationGroup}
  \bP_{ij}\,\bP_{jk}\,\bP_{ij}
  \ =\ \bP_{jk}\,\bP_{ij}\,\bP_{jk}\quad(i\neq k)
  \qquad\text{ and }\qquad
  \bP_{ij}^2\ =\ \bI\ \ .
\end{align}
  These relations are also depicted in Figure~\ref{fig:PG}.
  It is only slightly more cumbersome to work out the expression for
  the cyclic permutations on three sites since the latter can be
  expressed as a product of two transpositions,
  $\bT_{ijk}=\bP_{ij}\bP_{jk}$ and
  $\bT_{ijk}^{-1}=\bP_{jk}\bP_{ij}=\bT_{ijk}^\dag$. After some 
  straightforward algebra one then finds
\begin{align}
  \bT_{ijk}
  &\ =\ \bP_{ij}\bP_{jk}
   \ =\ -\tfrac{i}{2}\,f_{abc}\,S_i^aS_j^bS_k^c
       +\tfrac{1}{2}\,d_{abc}\,S_i^aS_j^bS_k^c
       +\tfrac{1}{N}\bigl[\vec{S}_i\cdot\vec{S}_j
       +\vec{S}_j\cdot\vec{S}_k+\vec{S}_i\cdot\vec{S}_k\bigr]
       +\tfrac{1}{N^2}\ \ .
\end{align}
  For our purposes we need to invert these relations and solve for the
  two cubic invariants involving the invariant rank-three tensors $f$
  and $d$. After some simple manipulations we find
\begin{subequations}
\begin{align}
  \label{eq:dSSS}
  d_{abc}\,S_i^aS_j^bS_k^c
  &\ =\ \bT_{ijk}+\bT_{ijk}^\dag
        -\tfrac{2}{N}\bigl[\bP_{ij}+\bP_{jk}+\bP_{ik}\bigr]
        +\tfrac{4}{N^2}\\[2mm]
  \label{eq:fSSS}
  f_{abc}\,S_i^aS_j^bS_k^c
  &\ =\ i(\bT_{ijk}-\bT_{ijk}^\dag)\ \ .
\end{align}
\end{subequations}

  The most convenient starting point for a replacement of the spin
  operators in terms of permutations seems to be
  \eqref{eq:HamiltonianSpinAlt}. After a lengthy but straightforward
  calculation one then recovers a Hamiltonian of the form
\begin{align}
  \label{eq:UniformHWithInvariants}
  H\ =\ g\,\bI+\sum_{i<j}(g_{ij}+g_{ji})\,\bP_{ij}
        +\sum_{i<j<k}\bigl(g_{ijk}\bT_{ijk}+\bar{g}_{ijk}\bT_{ijk}^\dag\bigr)\ \ ,
\end{align}
  where the individual constants are given by
\begin{subequations}
\begin{align}
  g
  &\ =\ \tfrac{(N-1)(N+2)}{2(N+1)}\sum_{i\neq j}|w_{ij}|^2
        -\tfrac{1}{2(N+1)}\sum_{i\neq j\neq k}\bar{w}_{ki}w_{kj}\\[2mm]
  g_{ij}
  &\ =\ \tfrac{N}{2}\,|w_{ij}|^2
        +\tfrac{1}{2}\sum_{k(\neq i,j)}\bar{w}_{ki}w_{kj}
        -\tfrac{1}{N+1}\,\text{Re}\Bigl[\bar{w}_{ji}
        \sum_{k(\neq j)}w_{jk}\Bigr]\\[2mm]
  g_{ijk}
  &\ =\ \tfrac{1}{2}\Bigl\{\bigl[\bar{w}_{ki}w_{kj}
        +\bar{w}_{jk}w_{ji}+\bar{w}_{ij}w_{ik}\bigr]
        -\tfrac{1}{(N+1)}\bigl[\bar{w}_{kj}w_{ki}
        +\bar{w}_{ji}w_{jk}+\bar{w}_{ik}w_{ij}
        \bigr]\Bigr\}\ \ .
\end{align}
\end{subequations}
  We believe that the expression \eqref{eq:UniformHWithInvariants},
  together with the decomposition of the Hilbert space as a
  representation of the symmetric group (employing the so-called
  Schur-Weyl duality), provides the computationally most efficient way
  of implementing the uniform spin model numerically, both in 1D and
  2D.

\subsection{The Hamiltonian for spins on a circle}
  
  We will now focus our attention to the 1D arrangement of spins on
  the unit circle with $w_{kl}=(z_k+z_l)/(z_k-z_l)$, see
  Section~\ref{sc:SetupCircle} for a concise definition of the setup.
  This choice implies a considerable number of non-trivial identities
  which allow us to simplify the Hamiltonian
  \eqref{eq:HamiltonianSpinAllOmega} and, in particular, to basically
  eliminate the three-spin couplings. First of all, one gets rid of
  complex conjugations in view of $\bar{w}_{kl}=-w_{kl}$. More
  importantly, the anti-symmetric three-spin coupling drops out due to
  $\Omega_{ijk}^A=0$. Finally, the symmetric three-spin coupling
  simplifies considerably due to $\Omega_{ijk}^T=2$. After employing
  these simplifications, the original Hamiltonian
  \eqref{eq:HamiltonianSpinAllOmega} may be rewritten as
\begin{align}
  H\ =\ -\tfrac{(N+2)(N-1)}{2N}\sum_{i\neq j}w_{ij}^2
        -\tfrac{N+2}{2(N+1)}\,H^{(2)}
        -\tfrac{N}{2(N+1)}\,H^{(3)}\ \ ,
\end{align}
  where we used the abbreviations
\begin{align}
  H^{(2)}
  \ =\ \sum_{i\neq j}
        \Bigl\{(N-1)w_{ij}^2
        +\sum_{k(\neq i,j)}w_{ki}w_{kj}
        \Bigr\}\,\vec{S}_i\cdot\vec{S}_j
  \qquad\text{ and }\qquad
  H^{(3)}
  \ =\ \sum_{i<j<k}d_{abc}\,S_i^aS_j^bS_k^c\ \ .
\end{align}
  In the next step we will consider the individual terms one by
  one. In order to simplify the quadratic term we shall use
  identity \eqref{eq:Sumww}. We then find
\begin{align}
  H^{(2)}
  &\ =\ \sum_{i\neq j}
        \Bigl\{L-2+(N+1)w_{ij}^2
        -w_{ij}(\xi_i-\xi_j)
        \Bigr\}\,\vec{S}_i\cdot\vec{S}_j\ \ .
\end{align}
  Finally, we employ \eqref{eq:Identitywz} and after a number of
  simplifications this leads to
\begin{align}
  H^{(2)}
  &\ =\ \sum_{i\neq j}
        \Bigl\{L-2+(N+1)+4(N+1)\tfrac{z_iz_j}{z_{ij}^2}
        -w_{ij}(\xi_i-\xi_j)
        \Bigr\}\,\vec{S}_i\cdot\vec{S}_j\\[2mm]
  &\ =\ \sum_{i\neq j}
        \Bigl\{4(N+1)\tfrac{z_iz_j}{z_{ij}^2}
        -w_{ij}(\xi_i-\xi_j)
        \Bigr\}\vec{S}_i\cdot\vec{S}_j
        +(L+N-1)\sum_{i\neq j}
        \vec{S}_i\cdot\vec{S}_j\ \ .
\end{align}
  The last part can be converted into an expression involving the
  total spin $\vec{S}=\sum_j\vec{S}_j$ using
\begin{align}
  \sum_{i\neq j}\vec{S}_i\cdot\vec{S}_j
  \ =\ \sum_{i,j}\vec{S}_i\cdot\vec{S}_j-\sum_j\vec{S}_j^2
  \ =\ \vec{S}^2-\tfrac{L(N^2-1)}{N}\ \ .
\end{align}
  Summing up all contributions we are left with
\begin{align}
  H^{(2)}
  &\ =\ 4(N+1)\sum_{i\neq j}
        \Bigl\{\tfrac{z_iz_j}{z_{ij}^2}
        -\tfrac{w_{ij}(\xi_i-\xi_j)}{4(N+1)}
        \Bigr\}\,\vec{S}_i\cdot\vec{S}_j
        +(L+N-1)\,\vec{S}^2
        -\tfrac{L(N^2-1)(L+N-1)}{N}\ \ .
\end{align}

  Next we turn our attention to the three-spin coupling. Our goal is
  to rewrite $H^{(3)}$ such that it again only involves the total spin
  $\vec{S}$. This can be achieved by restoring the summation over the
  full range of indices and enforcing the absence of the diagonal
  parts, and it leads to
\begin{equation}
\begin{split}
  H^{(3)}
  &\ =\ \tfrac{1}{6}\,d_{abc}\sum_{i,j,k}(1-\delta_{ij})
        (1-\delta_{ik}-\delta_{jk})\,S_i^aS_j^bS_k^c\\[2mm]
  &\ =\ \tfrac{1}{6}\,d_{abc}\sum_{i,j,k}\bigl[1
        -(\delta_{ij}+\delta_{ik}+\delta_{jk})
        +\delta_{ij}(\delta_{ik}+\delta_{jk})\bigr]
        \,S_i^aS_j^bS_k^c\ \ .
\end{split}
\end{equation}
  The individual contributions can be evaluated step by step,
  resulting first of all in
\begin{align}
  H_1^{(3)}
  &\ =\ \tfrac{1}{6}\,d_{abc}\sum_{i,j,k}S_i^aS_j^bS_k^c
   \ =\ \tfrac{1}{6}\,d_{abc}\,S^aS^bS^c
   \ =\ \tfrac{1}{6}\,\vec{S}^3\ \ .
\end{align}
  On the right hand side we defined the symbol $\vec{S}^3$ as
  the cubic invariant for the total spin which is obtained using the
  completely symmetric tensor. Then, splitting the summation into the
  diagonal part and the off-diagonal parts and using \eqref{eq:iik}
  one obtains
\begin{align}
  H_2^{(3)}
  \ =\ -\tfrac{N}{3}d_{abc}\sum_{i,j,k}
        (\delta_{ij}+\delta_{ik}+\delta_{jk})\,S_i^aS_j^bS_k^c
   \ =\ -(N^2-4)\,\vec{S}^2\ \ .
\end{align}
  Finally, using the Casimir eigenvalues $\vec{S}_i^2=(N^2-1)/N$ one
  recovers the expression
\begin{align}
  H_3^{(3)}
  \ =\ \tfrac{N}{3}d_{abc}\sum_{i,j,k}\delta_{ij}
       (\delta_{ik}+\delta_{jk})\,S_i^aS_j^bS_k^c
  \ =\ \tfrac{2L}{3N}(N^2-4)(N^2-1)\ \ .
\end{align}

  Putting all the previous calculations together and reordering the
  terms one obtains the Hamiltonian
\begin{equation}
\begin{split}
  \label{eq:modHaldaneShastry}
  H
  &\ =\ -2(N+2)\sum_{i\neq j}
        \Bigl\{\tfrac{z_iz_j}{z_{ij}^2}
        -\tfrac{w_{ij}(\xi_i-\xi_j)}{4(N+1)}
        \Bigr\}\vec{S}_i\cdot\vec{S}_j
        -\tfrac{(N+2)(N+2L)}{4(N+1)}\,\vec{S}^2
        -\tfrac{N}{12(N+1)}\,\vec{S}^3\\[2mm]
  &\qquad+\tfrac{L(N-1)(N+2)(3L+2N-1)}{6N}
        -\tfrac{(N+2)(N-1)}{2N}\sum_{i\neq j}w_{ij}^2\ \ .
\end{split}
\end{equation}
  Using relation \eqref{eq:HSzz}, we identify the first term as a
  modification of the $\SU(N)$ Haldane-Shastry model
  \cite{Haldane:1992PhRvL..69.2021H}.
  Since the coefficients
  $\xi_i$ vanish if the spins are positioned equidistantly on the unit
  circle, the model above includes the original Haldane-Shastry model
  as a special case.

  One can think of the Hamiltonian \eqref{eq:modHaldaneShastry} as a
  modification of the $\SU(N)$ Haldane-Shastry model. Its two-spin interaction
  has an altered (and actually rather intricate) distance dependence
  and the remaining terms correspond to the addition of two
  generalized chemical potentials. Indeed, while the usual chemical
  potential couples to the conserved particle number of a system of
  bosonic or fermionic particles, the coupling here favors spin
  configurations according to their conserved total Casimir
  eigenvalues. From this perspective, the Hamiltonian
  \eqref{eq:modHaldaneShastry} should be regarded as a special
  instance of the family
\begin{align}
  \label{eq:HChemicalPotential}
  H(\lambda_i)
  \ =\ H_{\text{mod\,HS}}+\lambda_2\,\vec{S}^2+\lambda_3\,\vec{S}^3
       +\cdots+\lambda_N\,\vec{S}^N\ \ .
\end{align}
  When writing this Hamiltonian we used that $\SU(N)$ has $N-1$
  independent Casimir operators which are described by symmetric
  tensors of rank $2,\ldots,N$. It should be noted that the additional
  terms in \eqref{eq:HChemicalPotential} turn a finite-size scaling
  analysis into a rather complicated issue, even if the first term
  $H_{\text{mod\,HS}}$ has a clean thermodynamic limit. In
  Section~\ref{sc:Loops} we will comment more on these subtleties.

\subsection{\label{sc:HamiltonianUniformED}The Hamiltonian for
  equidistant spins on a circle}

  In the case of an equidistant distribution of spins on the circle
  one has further simplifications such as (see \eqref{eq:uniformww})
\begin{align}
  \xi_i
  \ =\ 0
  \qquad\text{ and }\qquad
  \sum_{i\neq j}w_{ij}^2
  \ =\ -\tfrac{1}{3}L(L-1)(L-2)\ \ .
\end{align}
  In that case, the Hamiltonian essentially reduces to the
  Haldane-Shastry form and may be written as
\begin{align}
  \label{eq:modHaldaneShastryEquidistant}
  H
  &\ =\ (N+2)\sum_{k<l}
        \frac{\vec{S}_k\cdot\vec{S}_l}{\sin^2\tfrac{\pi}{L}(k-l)}
        -\tfrac{(N+2)(N+2L)}{4(N+1)}\,\vec{S}^2
        -\tfrac{N}{12(N+1)}\,\vec{S}^3
        +\tfrac{L(N+2)(N-1)(L^2+2N+1)}{6N}\ \ .
\end{align}
  This quantum spin system is exactly solvable since the underlying
  $\SU(N)$ Haldane-Shastry model can be treated analytically due to
  its Yangian symmetry
  \cite{Haldane:1992PhRvL..69.2021H,Ha:1993PhRvB..4712459H}. We note
  that the latter is not present and that originally degenerate
  multiplets are split when the chemical potentials $\lambda_2$ and
  $\lambda_3$ are added. However, this does not affect the statement
  that the model is exactly solvable. The thermodynamic limit of the
  first term in the Hamiltonian
  \eqref{eq:modHaldaneShastryEquidistant} is well known to be critical
  and described by a $\SU(N)_1$ WZW model
  \cite{Haldane:1992PhRvL..69.2021H,Bouwknegt:1996qv}, the starting
  point of our construction. The additional terms do not affect this
  conclusion but
  they modify the resulting WZW spectrum. Since the $\SU(N)$
  Haldane-Shastry model has already been studied thorougly in the past,
  we refrain from entering a more detailed discussion here.

\section{\label{sc:Mixed}Discussion of the mixed spin models}

  The structure of the Hamiltonians
  \eqref{eq:HamiltonianSpinAllOmegaGeneral} will be discussed for
  mixed spin models, involving both the fundamental and the
  anti-fundamental representation. Unfortunately, a reduction to
  two-spin couplings is not possible in this case, not even for an
  alternating chain of equidistant spins. However, we comment on
  possible simplifications in terms of generators of the walled Brauer
  algebra.

\subsection{Simplification of the Hamiltonian}

  In contrast to the discussion in Section~\ref{sc:Uniform} we now
  deal with the general situation where the physical spins may either
  transform in the fundamental or the anti-fundamental representation
  of $\SU(N)$. Accordingly, the set of sites $\bL$ splits into two
  subsets $\bS$ and $\dual{\bS}$ and there is a non-trivial grade map
  $d_\bullet:\bL\to\Integer_2$ encoding this decomposition (see our
  discussion around Eq.~\eqref{eq:GeneralHilbertSpace}).

  As in Section~\ref{sc:Uniform}, the two-spin interactions entering
  \eqref{eq:HamiltonianSpinAllOmegaGeneral} are still described in
  terms of $\SU(N)$-invariant operators. However, on a mixed Hilbert
  space $\cV\otimes\dual{\cV}$ the latter can of course not be
  implemented in terms of a permutation. Instead, the natural
  invariant operator (besides the identity) is the projection onto the
  singlet which, up to normalization, can be expressed in terms of the
  spin-spin coupling as
\begin{align}
  \bE_{ij}
  \ =\ -\vec{S}_i\cdot\vec{S}_j+\tfrac{1}{N}\ \ .
\end{align}
  The algebra of invariant operators is then generated by the
  operators $\bP_{ij}$ (for $i,j\in\bS$ or $i,j\in\dual{\bS}$) and
  $\bE_{ij}$ (for $i\in\bS$ and $j\in\dual{\bS}$ or vice versa). While
  the former obey the relations \eqref{eq:PermutationGroup} of the
  permutation group, the latter satisfy the Temperley-Lieb relations
\begin{align}
  \label{eq:TemperleyLieb}
  \bE_{ij}\,\bE_{jk}\,\bE_{ij}
  \ =\ \bE_{ij}
  \qquad\text{ and }\qquad
  \bE_{ij}^2
  \ =\ N\,\bE_{ij}
\end{align}
  with loop fugacity  $N=\dim(\cV)$.
  Of course, there are also non-trivial
  relations between the operators $\bP_{ij}$ and $\bE_{kl}$. If these
  relations are taken into account one is led to a representation of
  the so-called walled Brauer algebra, see Section~\ref{sc:Loops} for
  a more detailed explanation of this structure.

  Our ultimate goal is to rewrite the Hamiltonian
  \eqref{eq:HamiltonianSpinAllOmegaGeneral} in terms of invariant
  operators. It is obvious that the notation becomes too cumbersome
  when sticking to the symbols $\bP_{ij}$ and $\bE_{ij}$ since we
  always need to distinguish the different types of indices. Instead
  we will introduce a unified notation and define the invariant two
  site operator (for $i\neq j$)
\begin{align}
  \label{eq:DefQ}
  \bQ_{ij}
  \ =\ (-1)^{d_i+d_j}\,\vec{S}_i\cdot\vec{S}_j+\tfrac{1}{N}\ \ .
\end{align}
  We note that this operator is hermitean, i.e.\
  $\bQ_{ij}=\bQ_{ij}^\dag$. Moreover, it is symmetric in its indices,
  $\bQ_{ij}=\bQ_{ji}$. Depending on the nature of the indices of
  $\bQ_{ij}$ we either recover the usual permutation or the projection
  onto the singlet.

  Just as in Section~\ref{sc:Uniform}, our considerations easily
  generalize to three-spin interactions. As for the permutations, the
  invariant operators on three sites are either acting on two sites
  only or they are a product of two two-site operators. As it turns
  out, just two of these product operators are independent and they
  read
\begin{align}
  \label{eq:DefH}
  \bH_{ijk}
  \ =\ \bQ_{ij}\bQ_{jk}
  \qquad\text{ and }\qquad
  \bH_{ijk}^\dag
  \ =\ \bQ_{jk}\bQ_{ij}
  \ =\ \bH_{kji}\ \ .
\end{align}
  Permutations of the indices result in the same two operators but the
  precise outcome depends on the degree of all three labels
  involved. The defining relation \eqref{eq:DefQ} for the operators
  $\bQ$ in terms of the spin matrices imply the representation
\begin{equation}
\begin{split}
  \bH_{ijk}
  &\ =\  -\tfrac{i}{2}(-1)^{d_i+d_k}\,f_{abc}\,S_i^aS_j^bS_k^c
         -\tfrac{1}{2}(-1)^{d_i+d_j+d_k}\,d_{abc}\,S_i^aS_j^bS_k^c\\[2mm]
  &\qquad+\tfrac{1}{N}\bigl[(-1)^{d_i+d_j}\,\vec{S}_i\cdot\vec{S}_j
         +(-1)^{d_j+d_k}\,\vec{S}_j\cdot\vec{S}_k
         +(-1)^{d_i+d_k}\,\vec{S}_i\cdot\vec{S}_k\bigr]
         +\tfrac{1}{N^2}\ \ ,
\end{split}
\end{equation}
  which in turn allows to express the invariant operators in terms of
  spins. After some elementary algebra one finds the inversion
  formulas
\begin{subequations}
\begin{align}
  d_{abc}\,S_i^aS_j^bS_k^c
  &\ =\ -(-1)^{d_i+d_j+d_k}\Bigl[(\bH_{ijk}+\bH_{ijk}^\dag)
        -\tfrac{2}{N}\bigl[\bQ_{ij}
        +\bQ_{jk}+\bQ_{ik}\bigr]
        +\tfrac{4}{N^2}\Bigr]\\[2mm]
  f_{abc}\,S_i^aS_j^bS_k^c
  &\ =\ i(-1)^{d_i+d_k}(\bH_{ijk}-\bH_{ijk}^\dag)\ \ .
\end{align}
\end{subequations}
  We are now in a position to express the general Hamiltonian
  \eqref{eq:HamiltonianSpinAllOmegaGeneral} in terms of the invariant
  operators $\bQ$ and $\bH$.

  After a simple but lengthy computation we find the Hamiltonian
\begin{align}
  \label{eq:MixedHWithInvariants}
  H\ =\ g\,\bI+\sum_{i<j}(g_{ij}+g_{ji})\,\bQ_{ij}
        +\sum_{i<j<k}\bigl(g_{ijk}+g_{kij}+g_{jki}
        +g_{jik}+g_{kji}+g_{ikj}\bigr)\,\bH_{ijk}\ \ ,
\end{align}
  where the individual constants are given by
\begin{subequations}
  \label{eq:constants_H_mixed}
\begin{align}
  g
  &\ =\ \tfrac{N+2}{4(N+1)}\sum_{i\neq j}
        \Bigl\{2N-\Bigl[1+(-1)^{d_i+d_j}\Bigr]\Bigr\}|w_{ij}|^2
        -\tfrac{1}{2(N+1)}\sum_{i\neq j}(-1)^{d_i+d_j}
        \sum_{k(\neq i,j)}\bar{w}_{ki}w_{kj}\\[2mm]
  g_{ij}
  &\ =\ \tfrac{N}{4(N+1)}
        \Bigl[N+(N+2)(-1)^{d_i+d_j}\Bigr]|w_{ij}|^2
        +\tfrac{1}{2}(-1)^{d_i+d_j}\sum_{k(\neq i,j)}\bar{w}_{ki}w_{kj}\nonumber\\[2mm]
  &\qquad-\tfrac{1}{N+1}\,\text{Re}\Bigl[(-1)^{d_i}w_{ji}
        \sum_{k(\neq j)}(-1)^{d_k}\bar{w}_{jk}\Bigr]\\[2mm]
  g_{ijk}
  &\ =\ \tfrac{1}{4(N+1)}
        \Bigl\{\bar{w}_{ki}w_{kj}(-1)^{d_i}\Bigl[
        (N+2)(-1)^{d_k}+(-1)^{d_j}N\Bigr]\nonumber\\[2mm]
  &\qquad\qquad\qquad-\bar{w}_{ik}w_{ij}(-1)^{d_k}\Bigl[
        (N+2)(-1)^{d_i}-(-1)^{d_j}N\Bigr]
        \Bigr\}\ \ .
\end{align}
\end{subequations}
  It is possible to verify that this Hamiltonian reduces to
  \eqref{eq:UniformHWithInvariants} in the case of a uniform chain.

\subsection{\label{sc:HamiltonianAlternatingED}The Hamiltonian for
  equidistant spins on a circle}

  The Hamiltonian above may be simplified by assuming special
  positions for the spins. As before, the most convenient setup
  corresponds to equidistant spins on the circle, see
  Section~\ref{sc:SetupCircle}. Unlike in the uniform case, however,
  the three-spin couplings can not be simplified now since
  $\Omega_{ijk}^T$ fails to be constant (i.e.\ independent of the
  indices). Nevertheless using
  Eqs.~(\ref{eq:xi_and_minusw}-\ref{eq:uniformww}) the coupling
  constants \eqref{eq:constants_H_mixed} can be rewritten in the
  simplified form
\begin{subequations}
\begin{align}
  g
  &\ =\ 
  \tfrac{L(L-2)\bigl[L 
      (4 N^2+7N+2)-4(N^2+N-1)\bigr]}{24(N+1)} \\[2mm]
  g_{ij}
  &\ =\ -\tfrac{(-1)^{d_i+d_j}}{4(N+1)}
  \left[
    \left(N^2(1+(-1)^{d_i+d_j}) + 6 N + 4\right) w_{ij}^2 +
    2 (N + 1) (L - 2)
  \right]=g_{ji}\\[2mm]
  g_{ijk}
  &\ =\ \tfrac{1}{4(N+1)}
        \Bigl\{-w_{ki}w_{kj}(-1)^{d_i}\Bigl[
        (N+2)(-1)^{d_k}+(-1)^{d_j}N\Bigr]\nonumber\\[2mm]
  &\qquad\qquad\qquad+w_{ik}w_{ij}(-1)^{d_k}\Bigl[
        (N+2)(-1)^{d_i}-(-1)^{d_j}N\Bigr]
        \Bigr\}\ \ .
\end{align}
\end{subequations}
  The advantage of using these formulas is that they do not involve
  sums over the sites anymore, and thus can be efficiently evaluated
  numerically.

\section{\label{sc:Loops}Loop formulation and numerical
  implementation}

  As was discussed in Section~\ref{sc:Mixed}, the Hamiltonians for the
  mixed $\SU(N)$ spin models can be expressed in terms of generators of
  the walled Brauer algebra. We will now adopt a more abstract
  point of view and interpret the system from the perspective of loop
  models. This permits an efficient numerical implementation whose
  complexity is independent of the parameter $N$. For the alternating
  chain with equidistant spins on the unit circle we find evidence
  that the thermodynamic limit is described by a conformal field theory and
  we establish some properties of the latter.

\subsection{Definition of the loop model}

  Since the dimensions of the $\SU(N)$ representations $\cV$ and
  $\dual{\cV}$ are given by $N$, the dimension of the total Hilbert
  space \eqref{eq:GeneralHilbertSpace} grows as $\dim\cH=N^L$ where
  $L$ is the number of spins. Even for small values of $N\geq3$ the
  full implementation of the Hamiltonian quickly exceeds the available
  memory on computers. In order to avoid this complication we are
  seeking for a formulation of the diagonalization problem where the
  complexity is independent of $N$ but rather only depends on the
  number of spins $L$. This is achieved by relating our setup to loop
  models where $N=\dim\cV$ can be interpreted as the fugacity of the
  loops.

  In imaginary time, the exponential of the quantum spin Hamiltonian
  defines an evolution of the spin configuration along the
  longitudinal axis of a cylinder. It is convenient to interpret the
  $N$ different internal states of each spin as different types of
  particles. This allows one to illustrate the time evolution in terms
  of world-lines of these particles. Depending on whether the spin
  transforms in the fundamental or in the anti-fundamental
  representation we will either think of particles or their
  anti-particles and we will keep track of this difference by giving
  the corresponding world-lines opposite orientations. Let us now
  recall that the Hamiltonian \eqref{eq:MixedHWithInvariants} can be
  expressed in terms of either permutations or projections onto a
  singlet. In the world-line picture, these two operations correspond
  to the permutation of two (anti-)particles or to the pairwise
  annihilation of a particle and its associated anti-particle,
  followed by the creation of a mixed state involving all particle
  species. The latter process can be visualized diagrammatically by
  arcs connecting the strands horizontally, so that the particle type
  is conserved along a line. The operator $\bP_{ij}$ instead simply
  permutes the particles and admits a graphical representation as a
  crossing of the strands at sites  $i$ and $j$.
  During the dynamics loops may be formed, and each
  loop (contractible or not) carries a weight $\tr_\cV(\bI)=\dim\cV=N$. All
  processes just desribed must respect the orientation of
  world-lines.

  Given the previous correspondence, the study of the spin chain can
  now be approached graphically by studying the long-range model of
  crossing loops with weight $N$. We remark, however, that the
  transition from the spin chain to the loop formulation involves some
  subtleties. Indeed, for $N$ not sufficiently large (in a sense to be
  made precise below), some observables in the loop model may not have
  a counterpart in the spin chain.  This leads to the fact that
  (disregarding $\SU(N)$-related degeneracies) the spectrum of the
  loop model contains additional eigenvalues compared to that of the
  spin chain, as will be discussed in detail in
  Section~\ref{sec:spin_loop}. Although the geometrical loop
  formulation can be employed for a general setup with arbitrary
  positions of $\cV$ and $\dual{\cV}$, we will assume $L$ even in the
  following and focus on the alternating case
  $(\cV\otimes\dual{\cV})^{L/2}$.

\subsection{The walled Brauer algebra}

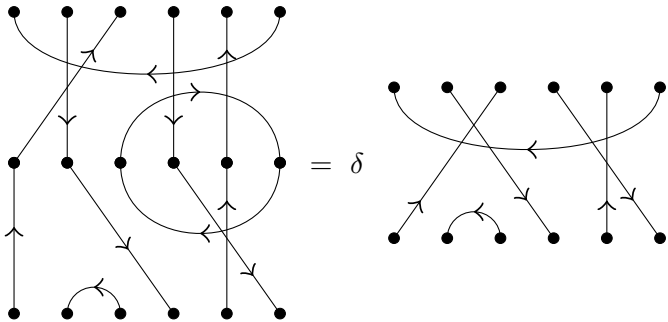
\begin{figure}[h]
\begin{center}
\begin{tikzpicture}[]
  \draw[postaction={decorate,decoration={ markings, mark=at position 
      0.575 with {\arrow[scale=2]{>}}}}] (.7*0,0) -- (.7*0,2);
  \draw[postaction={decorate,decoration={ markings, mark=at position 
      0.75 with {\arrow[scale=2]{>}}}}] (.7*4,0) -- (.7*4,2);
  \draw[postaction={decorate,decoration={ markings, mark=at position 
      0.575 with {\arrow[scale=2]{>}}}}] (.7*1,2) -- (.7*1+.7*2,0);
  \draw[postaction={decorate,decoration={ markings, mark=at position 
    0.75 with {\arrow[scale=2]{>}}}}] (.7*3,2) -- (.7*3+.7*2,0);
  \draw[postaction={decorate,decoration={ markings, mark=at position
      0.5 with {\arrow[scale=2]{>}}}}]
  (.7*2,0) arc (0:180:.7*.5);
 \draw[postaction={decorate,decoration={ markings, mark=at position
      0.5 with {\arrow[scale=2]{>}}}}]
  (.7*5,2) .. controls +(0,-1.25) and +(0,-1.25) .. (.7*2,2);
  \foreach \x in {0,1,2,3,4,5} {
    \draw[fill] (.7*\x,0) circle (2pt);
    \draw[fill] (.7*\x,2) circle (2pt);
  }
  \begin{scope}[yshift=2cm]
  \draw[postaction={decorate,decoration={ markings, mark=at position 
      0.75 with {\arrow[scale=2]{>}}}}] (.7*4,0) -- (.7*4,2);
  \draw[postaction={decorate,decoration={ markings, mark=at position 
      0.75 with {\arrow[scale=2]{>}}}}] (.7*0,0) -- (.7*2,2);
  \foreach \x in {1,3} {
    \draw[postaction={decorate,decoration={ markings, mark=at position 
      0.75 with {\arrow[scale=2]{>}}}}] (.7*\x,2) -- (.7*\x,0);
  }
  \draw[postaction={decorate,decoration={ markings, mark=at position
      0.5 with {\arrow[scale=2]{>}}}}]
  (.7*2,0) .. controls +(0,1.25) and +(0,1.25) .. (.7*5,0);
  \draw[postaction={decorate,decoration={ markings, mark=at position
      0.5 with {\arrow[scale=2]{>}}}}] 
  (.7*5,2) .. controls +(0,-1.1) and +(0,-1.1) .. (.7*0,2);
  \foreach \x in {0,1,2,3,4,5} {
    \draw[fill] (.7*\x,0) circle (2pt);
    \draw[fill] (.7*\x,2) circle (2pt);
  }
  \end{scope}
  \node at (4.25,2) {$=\ \delta$};
  \begin{scope}[yshift=1cm,xshift=5cm]
  \draw[postaction={decorate,decoration={ markings, mark=at position 
      0.25 with {\arrow[scale=2]{>}}}}] (.7*0,0) -- (.7*2,2);
  \draw[postaction={decorate,decoration={ markings, mark=at position 
      0.75 with {\arrow[scale=2]{>}}}}] (.7*1,2) -- (.7*3,0);
  \draw[postaction={decorate,decoration={ markings, mark=at position 
      0.75 with {\arrow[scale=2]{>}}}}] (.7*3,2) -- (.7*5,0);
  \draw[postaction={decorate,decoration={ markings, mark=at position 
      0.25 with {\arrow[scale=2]{>}}}}] (.7*4,0) -- (.7*4,2);
  \draw[postaction={decorate,decoration={ markings, mark=at position
      0.5 with {\arrow[scale=2]{>}}}}]
  (.7*2,0) arc (0:180:.7*.5);
  \draw[postaction={decorate,decoration={ markings, mark=at position
      0.5 with {\arrow[scale=2]{>}}}}] 
  (.7*5,2) .. controls +(0,-1.1) and +(0,-1.1) .. (.7*0,2);
  \foreach \x in {0,1,2,3,4,5} {
    \draw[fill] (.7*\x,0) circle (2pt);
    \draw[fill] (.7*\x,2) circle (2pt);
  }
  \end{scope}
\end{tikzpicture}
  \caption{\label{fig:multi}An example for the multiplication of
    diagrams.}
\end{center}
\end{figure}

  We consider now the loop model and discuss abstractly the properties
  of the algebra of diagrams associated to the elementary interactions
  $\bE_{ij}$ and $\bP_{ij}$. A diagram is a set of $L$ top nodes and
  $L$ bottom nodes, numbered from left to right, so that each node is
  connected to precisely one other by a line.  We call the lines
  connecting bottom to top nodes ``through lines''. The diagrams
  relevant for our analysis have some constraints. As before we assign
  alternating orientations to the lines, and consider only diagrams
  whose connectivities respect the orientation.  The linear span over
  $\Complex$ of these diagrams is turned into an algebra by specifying
  a product $D_1\cdot D_2$, which is given by the diagram obtained by
  placing $D_1$ on top of $D_2$ and replacing all loops formed with a
  fixed weight $\delta\in\Complex$. In Figure~\ref{fig:multi} this
  multiplication law for diagrams is illustrated in a specific
  example.  The algebra so formed is called the walled Brauer algebra
  $\wBr(\delta)$.  Clearly the relation with the spin chains we would
  like to study comes about when we specify $\delta=N$, but it is
  useful to regard $\delta$ as an arbitrary complex number for the
  moment.

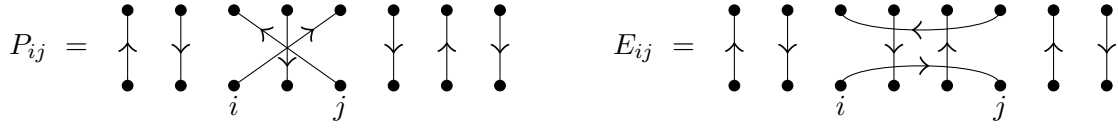
\begin{figure}
\begin{center}
\begin{tikzpicture}[]
  \draw (0,.5) node[left] {$P_{ij}\ =\quad$};
  \foreach \x in {0,6} {
    \draw[postaction={decorate,decoration={ markings, mark=at position 
    0.575 with {\arrow[scale=2]{>}}}}] (.7*\x,0) -- (.7*\x,1);
  }
  \foreach \x in {1,5,7} {
    \draw[postaction={decorate,decoration={ markings, mark=at position 
    0.575 with {\arrow[scale=2]{>}}}}] (.7*\x,1) -- (.7*\x,0);
  }
  \draw[postaction={decorate,decoration={ markings, mark=at position 
      0.75 with {\arrow[scale=2]{>}}}}] (.7*3,1) -- (.7*3,0);
  \draw (.7*2,0) node[below] {$i$};
  \draw (.7*4,0) node[below] {$j$};
  \draw[postaction={decorate,decoration={ markings, mark=at position
      0.75 with {\arrow[scale=2]{>}}}}] (.7*2,0) -- (.7*4,1);
  \draw[postaction={decorate,decoration={ markings, mark=at position
      0.75 with {\arrow[scale=2]{>}}}}] (.7*4,0) -- (.7*2,1);
  \foreach \x in {0,1,2,3,4,5,6,7} {
    \draw[fill] (.7*\x,0) circle (2pt);
    \draw[fill] (.7*\x,1) circle (2pt);
  }
\end{tikzpicture}
  \qquad\quad
\begin{tikzpicture}
  \draw (0,.5) node[left] {$E_{ij}\ =\quad$};
  \foreach \x in {0,4,6} {
    \draw[postaction={decorate,decoration={ markings, mark=at position 
    0.575 with {\arrow[scale=2]{>}}}}] (.7*\x,0) -- (.7*\x,1);
  }
  \foreach \x in {1,3,7} {
    \draw[postaction={decorate,decoration={ markings, mark=at position 
    0.575 with {\arrow[scale=2]{>}}}}] (.7*\x,1) -- (.7*\x,0);
  }
  \draw (.7*2,0) node[below] {$i$};
  \draw (.7*5,0) node[below] {$j$};
  \draw[postaction={decorate,decoration={ markings, mark=at position 
    0.55 with {\arrow[scale=2]{>}}}}] (.7*2,0) .. controls +(0,.35) and +(0,.35) .. (.7*5,0);
  \draw[postaction={decorate,decoration={ markings, mark=at position 
    0.55 with {\arrow[scale=2]{>}}}}] (.7*5,1) .. controls +(0,-.35) and +(0,-.35) .. (.7*2,1);
  \foreach \x in {0,1,2,3,4,5,6,7} {
    \draw[fill] (.7*\x,0) circle (2pt);
    \draw[fill] (.7*\x,1) circle (2pt);
  }
\end{tikzpicture}
  \caption{\label{fig:EP}The elements $P_{ij}$ and $E_{ij}$ of the
    walled Brauer algebra.}
\end{center}
\end{figure}

  We now summarize some properties of $\wBr(\delta)$ that we need
  below for the discussion of the spectrum of the spin chain.  We
  denote by $\Ed_{ij},\Pd_{ij}$ the abstract diagrams corresponding to
  the action of $\bE_{ij},\bP_{ij}$ in the loop formulation of the
  spin model, see Figure~\ref{fig:EP}.  

\begin{figure}
\begin{center}
\begin{tikzpicture}
\begin{scope}
  \foreach \x in {0,1,2,3.5,4.5,5.5} {
    \draw[fill] (.7*\x,0) circle (2pt);
    \draw[fill] (.7*\x,3) circle (2pt);
  }
  \draw (.7*0,0) -- (.7*2,2) -- (.7*2,3);
  \draw (.7*1,0) -- (.7*0,1) -- (.7*0,2) -- (.7*1,3);
  \draw (.7*2,0) -- (.7*2,1) -- (.7*0,3);
  \draw (.7*2.75,1.5) node {$=$};
  \draw (.7*3.5,0) -- (.7*3.5,1) -- (.7*5.5,3);
  \draw (.7*4.5,0) -- (.7*5.5,1) -- (.7*5.5,2) -- (.7*4.5,3);
  \draw (.7*5.5,0) -- (.7*3.5,2) -- (.7*3.5,3);
  \draw (.7*0,0) node[below] {$i$};
  \draw (.7*1,0) node[below] {$j$};
  \draw (.7*2,0) node[below] {$k$};
  \draw (.7*3.5,0) node[below] {$i$};
  \draw (.7*4.5,0) node[below] {$j$};
  \draw (.7*5.5,0) node[below] {$k$};
\end{scope}
\begin{scope}[xshift=6cm,yshift=.5cm]
  \foreach \x in {0,1,2.5,3.5} {
    \draw[fill] (.7*\x,0) circle (2pt);
    \draw[fill] (.7*\x,2) circle (2pt);
  }
  \draw (.7*0,0) -- (.7*1,1) -- (.7*0,2);
  \draw (.7*1,0) -- (.7*0,1) -- (.7*1,2);
  \draw (.7*1.75,1) node {$=$};
  \draw (.7*2.5,0) -- (.7*2.5,2);
  \draw (.7*3.5,0) -- (.7*3.5,2);
  \draw (.7*0,0) node[below] {$i$};
  \draw (.7*1,0) node[below] {$j$};
  \draw (.7*2.5,0) node[below] {$i$};
  \draw (.7*3.5,0) node[below] {$j$};
\end{scope}
\end{tikzpicture}
\end{center}
  \caption{\label{fig:PG}The relations \eqref{eq:PermutationGroup} of
    the permutation group in the loop formulation.
  The lines are assumed to have the same orientation.}
\end{figure}
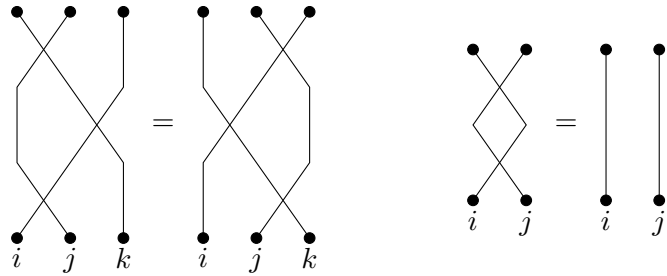

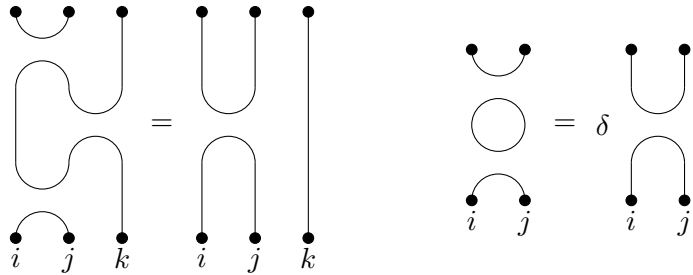
\begin{figure}
\begin{center}
\begin{tikzpicture}
\begin{scope}
  \foreach \x in {0,1,2,3.5,4.5,5.5} {
    \draw[fill] (.7*\x,0) circle (2pt);
    \draw[fill] (.7*\x,3) circle (2pt);
  }
  \draw (.7*2,0) -- (.7*2,1);
  \draw (.7*1,0) arc (0:180:.7*.5);
  \draw (.7*1,1) arc (360:180:.7*.5);
  \draw (0,1) -- (0,2);
  \draw (.7*2,1) arc (0:180:.7*.5);
  \draw (.7*2,2) arc (360:180:.7*.5);
  \draw (.7*2,2) -- (.7*2,3);
  \draw (.7*1,2) arc (0:180:.7*.5);
  \draw (.7*1,3) arc (360:180:.7*.5);
  \draw (.7*2.75,1.5) node {$=$};
  \draw (.7*5.5,0) -- (.7*5.5,3);
  \draw (.7*3.5,0) -- (.7*3.5,1);
  \draw (.7*3.5,2) -- (.7*3.5,3);
  \draw (.7*4.5,0) -- (.7*4.5,1);
  \draw (.7*4.5,2) -- (.7*4.5,3);
  \draw (.7*4.5,1) arc (0:180:.7*.5);
  \draw (.7*4.5,2) arc (360:180:.7*.5);
  \draw (.7*0,0) node[below] {$i$};
  \draw (.7*1,0) node[below] {$j$};
  \draw (.7*2,0) node[below] {$k$};
  \draw (.7*3.5,0) node[below] {$i$};
  \draw (.7*4.5,0) node[below] {$j$};
  \draw (.7*5.5,0) node[below] {$k$};
\end{scope}
\begin{scope}[xshift=6cm,yshift=.5cm]
  \foreach \x in {0,1,3,4} {
    \draw[fill] (.7*\x,0) circle (2pt);
    \draw[fill] (.7*\x,2) circle (2pt);
  }
  \draw (.7*1,0) arc (0:180:.7*.5);
  \draw (.7*1,1) arc (360:180:.7*.5);
  \draw (.7*1,1) arc (0:180:.7*.5);
  \draw (.7*1,2) arc (360:180:.7*.5);
  \draw (.7*1.75,1) node {$=$} +(.5,0) node {$\delta$};
  \draw (.7*3,0) -- (.7*3,.5);
  \draw (.7*4,0) -- (.7*4,.5);
  \draw (.7*3,1.5) -- (.7*3,2);
  \draw (.7*4,1.5) -- (.7*4,2);
  \draw (.7*4,.5) arc (0:180:.7*.5);
  \draw (.7*4,1.5) arc (360:180:.7*.5);
  \draw (.7*0,0) node[below] {$i$};
  \draw (.7*1,0) node[below] {$j$};
  \draw (.7*3,0) node[below] {$i$};
  \draw (.7*4,0) node[below] {$j$};
\end{scope}
\end{tikzpicture}
\end{center}
  \caption{\label{fig:TL}The relations \eqref{eq:TemperleyLieb} in the
    loop formulation. In the $\SU(N)$ spin models, the loop fugacity
    $\delta$ is given by $N=\dim\cV$.
    The lines are assumed to have alternating orientations.
  }
\end{figure}

  The walled Brauer algebra $\wBr(\delta)$ can be presented as a set
  of generators and relations. As generators it is sufficient to take
  the permutations $\Pd_{i,i+2}$ with $i=1,\dots,L-2$ together with
  $\Ed_{12}$. On products of these generators one then imposes the
  natural relations which ensure that diagrams with the same
  connectivity are identified and that loops 
  have weight $\delta$. (Note that the remaining elements $\Ed_{ij}$
  can be obtained by multiplying $\Ed_{12}$ from the left and the
  right by the permutation exchanging $(1,2)$ with either $(i,j)$ or $(j,i)$,
  depending on their parity.) The diagrammatic form of the relations
  \eqref{eq:PermutationGroup} and \eqref{eq:TemperleyLieb} is depicted
  in Figures~\ref{fig:PG} and~\ref{fig:TL}.

  We stress furthermore that by flipping the arrows on all the odd
  nodes one obtains diagrams belonging to (the group algebra of) the
  symmetric group $\symm_{L}$. In particular this shows that the
  dimension of $\wBr(\delta)$ equals that of $\symm_{L}$, namely $L!$,
  independently of $\delta$.

  In the loop formulation the Hamiltonian is expressed in terms of
  diagrams of the walled Brauer algebra. We now discuss the problem of
  diagonalizing such an operator. It will be convenient to reduce the
  dimension of the space of states of our problem by looking at
  sub-sectors labeled by some quantum numbers determining individual
  representations of the algebra at hand. The walled Brauer algebra is
  a finite dimensional algebra and as such all its irreducible
  representations can be realized by acting with the algebra on itself
  (this is called the regular representation). This means that we can
  restrict ourselves to studying the action of the walled Brauer
  algebra on diagrams. Our next goal is to find subspaces on which
  this action is closed. For this purpose we introduce the notation
  $D=X_{v,w,\sigma}$ for a diagram $D$, where $v$ is the configuration
  of the $(L-K)/2$ northern arcs, $w$ that of the $(L-K)/2$ bottom
  arcs, and $\sigma\in \symm_{K/2}\times \symm_{K/2}\subset\symm_K$
  (the two factors refer to the two orientations) is a permutation
  specifying how the nodes not occupied by arcs are connected:
  $\sigma(i)=j$ indicates that the $i^{\text{th}}$ bottom node is
  connected to node $j$.

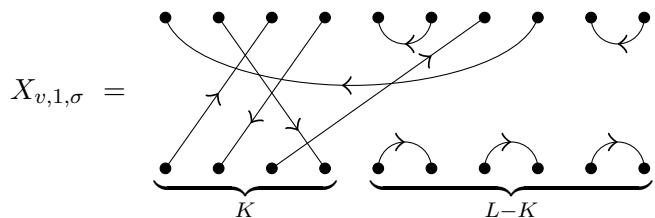
\begin{figure}[h]
\begin{center}
\begin{tikzpicture}[]
  \draw (0,1) node[left] {$X_{v,1,\sigma}\ =\quad$};
  \draw[postaction={decorate,decoration={ markings, mark=at position 
      0.75 with {\arrow[scale=2]{>}}}}] (.7*1,2) -- (.7*3,0);
  \draw[postaction={decorate,decoration={ markings, mark=at position 
      0.7 with {\arrow[scale=2]{>}}}}] (.7*3,2) -- (.7*1,0);
  \draw[postaction={decorate,decoration={ markings, mark=at position
      0.5 with {\arrow[scale=2]{>}}}}] (.7*0,0) -- (.7*2,2);
  \draw[postaction={decorate,decoration={ markings, mark=at position
      0.75 with {\arrow[scale=2]{>}}}}] (.7*2,0) -- (.7*6,2);
  \draw[postaction={decorate,decoration={ markings, mark=at position
      0.5 with {\arrow[scale=2]{>}}}}] (.7*4,0) arc (180:0:.7*.5);
  \draw[postaction={decorate,decoration={ markings, mark=at position
      0.5 with {\arrow[scale=2]{>}}}}] (.7*6,0) arc (180:0:.7*.5);
  \draw[postaction={decorate,decoration={ markings, mark=at position
      0.5 with {\arrow[scale=2]{>}}}}] (.7*8,0) arc (180:0:.7*.5);
  \draw[postaction={decorate,decoration={ markings, mark=at position
      0.5 with {\arrow[scale=2]{<}}}}] (.7*0,2)
    .. controls +(0.2,-.6) and +(-1,0) .. (.7*3.5,1.1)
    .. controls +(1,0) and +(-0.2,-.6) .. (.7*7,2);
  \draw[postaction={decorate,decoration={ markings, mark=at position
      0.5 with {\arrow[scale=2]{>}}}}] (.7*5,2) arc (0:-180:.7*.5);
  \draw[postaction={decorate,decoration={ markings, mark=at position
      0.5 with {\arrow[scale=2]{>}}}}] (.7*9,2) arc (0:-180:.7*.5);
  \foreach \x in {0,1,2,3,4,5,6,7,8,9} {
    \draw[fill] (.7*\x,0) circle (2pt);
    \draw[fill] (.7*\x,2) circle (2pt);
  }
  \draw (.7*1.5,0) node[below] {$\underbrace{\hspace{24mm}}_{K}$};
  \draw (.7*6.5,0) node[below] {$\underbrace{\hspace{37mm}}_{L-K}$};
\end{tikzpicture}
\caption{\label{fig:bott_nodes}Illustration of an element
  $X_{v,1,\sigma}$.  The index $1$ stands for the configuration chosen
  at the bottom, $v$ corresponds to the configuration of arcs at the
  top together with the choice of nodes where the arcs lie, and
  $\sigma$ encodes the permutation of through lines.}
\end{center}
\end{figure}

  One can easily convince oneself that the number of through lines can
  only be lowered under the action of the algebra but never increased.
  It is then reasonable to work in a basis of diagrams which is
  ordered in an increasing fashion according to the number of through
  lines. In such a basis, any Hamiltonian based on the walled Brauer
  algebra will have a block upper-triangular structure.  To compute
  the eigenvalues of the Hamiltonian it is then sufficient to restrict
  to the blocks by acting on diagrams with a {\em fixed} number of
  through lines. This reduces the calculational effort and can be
  implemented in practice by setting the action on a state to zero if
  the number of through lines is reduced. Furthermore, the
  multiplication rule of the walled Brauer algebra implies that the
  action of the Hamiltonian on a given diagram modifies only the
  connectivity of its top row of nodes.  As a consequence, the
  eigenvalues have a huge degeneracy. The latter can be removed by
  restricting oneself to diagrams where the configurations on the
  bottom are frozen to a given one.\footnote{The degeneracy results
    from the fact that the Brauer algebra admits an action by left
    multiplication and another one by right multiplication. Our way of
    concatenating diagrams in the regular representation singles out
    the left multiplication. The (irrelevant) right multiplication may
    then be used to freeze the bottom configuration.}  For
  definiteness, fixed the number $K$ of through lines (which is always
  even for $L$ even), we choose the leftmost $K$ nodes at the bottom
  to be connected with the top, and the remaining $L-K$ nodes to have
  the arc configuration connecting node $K+i$ with $K+i+1$  for
  $i=0,2,\dots, L-K-2$.  Using the notation introduced above,
  such elements are denoted by $X_{v,1,\sigma}$ where $1$ refers to
  the fixed configuration of bottom arcs chosen. For a fixed arc
  connectivity $v$ with $(L-K)/2$ arcs, the diagrams $X_{v,1,\sigma}$
  differ by the permutation $\sigma\in \symm_{K/2} \times \symm_{K/2}$
  encoding how the through lines with the same orientation are
  permuted, see Figure~\ref{fig:bott_nodes} for an example.

  The Hamiltonian can be further block diagonalized by projecting onto
  subspaces which transform according to irreducible representations
  of $\symm_{K/2}\times \symm_{K/2}$.  We recall that the irreducible
  representations $S^\lambda$ of $\symm_m$, the so-called Specht
  modules \cite{Sagan2001}, are labeled by partitions of $m$, denoted
  by $\lambda \vdash m$. A partition $\lambda\vdash m$ is a
  non-increasing sequence of positive integers which sum up to $m$:
  $\lambda=(\lambda_1,\dots,\lambda_{\ell(\lambda)})$, with
  $\lambda_1\ge \lambda_2\ge \cdots \ge \lambda_{\ell(\lambda)}\ge1$
  and $\lambda_1+\cdots+\lambda_{\ell(\lambda)}=m$.  $\ell(\lambda)$
  is called the length of the partition. A convenient way of depicting
  the partition $\lambda$ is in terms of Young tableaux. In our case,
  we have two identical copies of the permutation group and hence the
  irreducible representations $S^{\mu,\nu}=S^\mu\times S^\nu$ of
  $\symm_{K/2}\times \symm_{K/2}$ are indexed by a pair of partitions
  $(\mu,\nu)\vdash (K/2,K/2)$. The resulting representation of the
  walled Brauer algebra will be denoted by $\mathcal{W}_L(\mu,\nu)$.

  We now present an explicit construction of the space
  $\mathcal{W}_L(\mu,\nu)$ in terms of a suitable projection on the
  set of all diagrams \cite{Cox2008}. Denote by $I_{L}^K$ the space
  spanned by the diagrams $X_{v,1,\id}$ where $v$ is any allowed arc
  configuration of the top row with \emph{exactly} $(L-K)/2$ arcs and
  $\id$ is the identity permutation (no crossings between through
  lines). $\mathcal{W}_L(\mu,\nu)$ is given by the tensor product of
  $I_{L}^K$ with $S^{\mu,\nu}$, and its basis elements are of the form
  $X_{v,1,\id}\otimes x $, where $x$ runs through a basis of
  $S^{\mu,\nu}$. (The construction of a basis of the irreducible
  representation of the symmetric group is standard, see
  e.g.~\cite{Sagan2001}. Since it will not be explicitly needed later
  on, we omit the details here.)  The action of a diagram on this
  basis is given by concatenation from above on $X_{v,1,\id}$.  The
  result is set to zero if the number of through lines is reduced,
  since this would modify the pattern at the bottom
  nodes. Furthermore, a permutation $\sigma\in \symm_{K/2}\times
  \symm_{K/2}$ of the through lines would also produce an element
  outside of $I_L^K$ since the permutation $\id$ is replaced by
  something else. However, such a permutation of through lines will be
  absorbed by acting on the irreducible representation $S^{\mu,\nu}$
  instead.

  The spaces $\mathcal{W}_L(\mu,\nu)$ with $(\mu,\nu)\vdash(K/2,K/2)$
  and $K=0,2,\dots,L$ are the essential building blocks on which we
  want to diagonalize our Hamiltonian.  Note that the dimension of
  $I_{L}^K$ corresponds to all possible ways of choosing arc
  configurations at the top nodes with exactly $(L-K)/2$
  arcs. It is determined by $\binom{L/2}{(L-K)/2} \binom{L/2}{(L-K)/2}
  ((L-K)/2)!$, where the last factor comes from the possible
  ways of pairing $(L-K)/2$ objects with $(L-K)/2$ other
  objects. Then due to the tensor product structure we have
\begin{equation}
\begin{split}
  d_{L}^{\mu,\nu}&:=\dim\bigl(\mathcal{W}_L(\mu,\nu)\bigr)
  =
  \binom{L/2}{(L-K)/2}^2 
  \left((L-K)/2\right)!\,\dim(S^\mu) \dim(S^\nu)\\
  &=
  \frac{\left(L/2\right)!\left(L/2\right)!}
  {\left(K/2\right)!\left(K/2\right)!
    \left((L-K)/2\right)!}\,\dim(S^\mu) \dim(S^\nu)
  \, .
\end{split}
\end{equation}
  In particular, one obtains the dimension $(L/2)!$ for $K=0$ and
  $L(L/2)!/2$ for $K=2$.

  We now briefly comment on the properties of the representations
  $\mathcal{W}_L(\mu,\nu)$. It has been proven in
  \cite[Thm.~6.3]{Cox2008} that $\wBr(\delta)$ is semisimple when
  $\delta\not\in\Integer$ or $|\delta|\ge L-1$. 
  For these values of
  $\delta$ all representations are fully reducible and the
  representations $\mathcal{W}_L(\mu,\nu)$ form a complete set of
  irreducible representations. Moreover, denoting the matrix algebra
  of $d\times d$ matrices (over $\Complex$) by $\mathcal{M}_{d}$, one
  has the decomposition (as algebras and bimodules)
\begin{align}
  \label{eq:wBr_dec}
  \wBr(\delta)\cong \bigoplus_{K=0,2}^{L}
  \bigoplus_{\mu,\nu\,\vdash K/2}
  \mathcal{M}_{d_{L}^{\mu,\nu}}\, ,
\end{align}
  where the subscript indicates that the first summation runs in steps
  of two. It is useful to think of the $\mathcal{M}_{d_{L}^{\mu,\nu}}$
  as the space of linear maps on $\mathcal{W}_L(\mu,\nu)$. The
  decomposition \eqref{eq:wBr_dec} is supported by the following
  comparison of dimensions:
\begin{align}
  \sum_{K=0,2}^L 
  \sum_{\mu,\nu\,\vdash K/2}\dim(M_{d_{L}^{\mu,\nu}})
  &=
  \sum_{K=0,2}^L \sum_{\mu,\nu\,\vdash K/2} 
  \left(
  \frac{\left(L/2\right)!\left(L/2\right)!}
  {\left(K/2\right)!\left(K/2\right)!
    \left((L-K)/2\right)!}
  \dim(S^\mu) \dim(S^\nu)
  \right)^2\nonumber\\
  &=
  \sum_{K=0,2}^L
  \left(
  \frac{\left(L/2\right)!\left(L/2\right)!}
  {\left(K/2\right)!\left(K/2\right)!
    \left((L-K)/2\right)!}
  \right)^2
  \Biggl(\,
  \sum_{\mu\,\vdash K/2}
  \bigl(
  \dim(S^\mu) 
  \bigr)^2
  \Biggr)^2
  \\
  &=
  \sum_{K=0,2}^L
  \binom{L/2}{K/2}^2
  \bigl((L/2)!\bigr)^2
  =L!
  \, ,\nonumber
\end{align}
  where for the third equality we used that $m! = \dim(\symm_m)=
  \sum_{\mu\,\vdash m}\bigl(\dim(S^\mu)\bigr)^2$. If $\delta$ instead
  is a small integer, Eq.~\eqref{eq:wBr_dec} ceases to be true (as a
  decomposition of algebras and bimodules) and the representation
  theory of $\wBr(\delta)$ gets much more complicated. For the purpose
  of computing the eigenvalues of the Hamiltonian, entering such
  details is not necessary, and in the following we will simply
  restrict the numerical diagonalization to the spaces
  $\mathcal{W}_L(\mu,\nu)$.

  We conclude this section by summarizing what we have done and what
  we have gained. The spin chains have been mapped onto a geometrical
  model of crossing loops with long-range interactions. The number $N$
  of the $\SU(N)$ spin chains enters in the loop model as a
  parameter. Since the dimension of the space of states
  $\mathcal{W}_L(\mu,\nu)$ does not depend on $N$, our reformulation
  allows us to efficiently investigate the $\SU(N)$ spin chains for
  $N$ arbitrary large, a task which is not feasible when diagonalizing
  the spin chain directly. Since our presentation silently skipped
  over a few subtleties, we will devote the next section to a precise
  discussion of how to reconstruct the spectrum of the spin chain from
  that of the loop model.

\subsection{\label{sec:spin_loop}Relation with the spin chains}

  So far we have discussed the motivation for a loop reformulation of
  our spin chains and reviewed some of the algebraic properties of the
  loop model. In this section, we will now comment on the precise
  relation between the energy spectrum in the loop model as compared
  to that of the spin chain. The algebraic considerations which follow
  are based on \cite{Benkart1994}.

  In the following it will be convenient to view our models from the
  perspective of $\GL(N)$ instead of $\SU(N)$. This is justified since
  the Hamiltonian for the alternating chain commutes with the
  generators of $\GL(N)$ which span the Lie algebra $\gl_N$. The
  Hamiltonian can be regarded as an element of the centralizer algebra
  $Z(\gl_N)$, the algebra of all linear operators on
  $\bigl(\cV\otimes\dual{\cV}\bigr)^{\otimes L/2}$ that
  commute with the action of $\gl_N$. As we shall discuss below, the
  algebra $Z(\gl_N)$ is closely related to the walled Brauer
  algebra. In fact, in the ``stable'' (but rather unphysical) regime
  where $N\geq L$ one has $Z(\gl_N)\cong\wBr(N)$ and a Hilbert space
  decomposition of the form \cite{Benkart1994} \footnote{We stress
    that the bound for the equivalence of $Z(\gl_N)$ and $\wBr(N)$
    is different from $N\ge L-1$, the range for the semisimplicity 
    of $\wBr(N)$ mentioned above \cite[Thm.~6.3]{Cox2008}.
  }
\begin{align}
  \label{eq:VVstar_gl_Br}
  \bigl(\cV\otimes\dual{\cV}\bigr)^{\otimes L/2}\bigr|_{\gl_N\otimes\wBr(N)}
  \ \cong\ \bigoplus_{K=0,2}^L\bigoplus_{\mu,\nu\,\vdash K/2}
           V\bigl([\mu,\nu]_N\bigr)\otimes\mathcal{W}_L^{\mu,\nu}
  \qquad(\text{for }N\geq L)\, .
\end{align}
  The symbol $V\bigl([\mu,\nu]_N\bigr)$ refers to the
  $\gl_N$-representation corresponding to the highest weight
  $[\mu,\nu]_N$. It is obtained from the two partitions
  $\mu=(\mu_1,\dots,\mu_{\ell(\mu)})$ and
  $\nu=(\nu_1,\dots,\nu_{\ell(\nu)})$ by setting\footnote{Here we chose the
    $\gl_N$ Cartan subalgebra as the diagonal matrices with a single
    unit element. The corresponding weight of $SU(N)$ is therefore
    given by $(\mu_1-\mu_2,\mu_2-\mu_3,\ldots)$.}
\begin{align}
  \label{eq:glweight}
  [\mu,\nu]_N
  \ :=\ \bigl[\mu_1,\mu_2,\dots,\mu_{\ell(\mu)},
  0^{N-\ell(\mu)-\ell(\nu)},
  -\nu_{\ell(\nu)},-\nu_{\ell(\nu)-1},\dots,-\nu_1\bigr]
  \, .
\end{align}
  The condition $N\geq L$ ensures that $\ell(\mu)+\ell(\nu)\le N$
  and that this assignment is well defined.

  For the task of finding the spin chain spectrum we now focus on
  the action of $\wBr(N)$. Denoting by $\tilde{z}_{\mu,\nu}=\dim
  V\bigl([\mu,\nu]_N\bigr)$ the degeneracy associated with the
  $\gl_N$-symmetry, the relevant information in the decomposition
  \eqref{eq:VVstar_gl_Br} is
\begin{align}
  \label{eq:VVstar_wBr}
  (\cV\otimes\dual{\cV})^{\otimes L/2}\big|_{\wBr(N)}
  \ \cong\ \bigoplus_{K=0,2}^L\,\bigoplus_{\mu,\nu\,\vdash K/2}
  \tilde{z}_{\mu,\nu}\,\mathcal{W}_L^{\mu,\nu}
  \qquad(\text{for }N\geq L)\, .
\end{align}
  Indeed, the Hamiltonian can be regarded as an element of $\wBr(N)$
  and hence it is sufficient to consider the diagonalization problem
  on the invariant subspaces $\mathcal{W}_L^{\mu,\nu}$ entering the
  decomposition \eqref{eq:VVstar_wBr}. This establishes the connection
  to the loop model. It is then obvious how the spectrum of the spin
  chain can be reconstructed from that of the loop model, at least as
  long as $N\geq L$.

  The situation is more complicated in the regime $N<L$ which is
  relevant for the thermodynamic limit of the chain. In this case we
  need to understand the decomposition of the Hilbert space
  $(\cV\otimes\dual{\cV})^{\otimes L/2}$ with respect to $\gl_N\otimes
  Z(\gl_N)$. Let us introduce a map $\phi$ which represents the action
  of the walled Brauer algebra $\wBr(N)$ on the Hilbert space of the
  spin chain. This map $\phi$ sends $\Ed_{ij}$ to $\bE_{ij}$ and
  $\Pd_{ij}$ to $\bP_{ij}$ and it constitutes an algebra homomorphism,
  i.e.\ it satisfies $\phi(D_1D_2)=\phi(D_1)\phi(D_2)$ for all
  $D_1,D_2\in\wBr(N)$. As already observed, $\bE_{ij}$ and $\bP_{ij}$
  commute with $\gl_N$, so that $\phi$ can be regarded as a map from
  $\wBr(N)$ to $Z(\gl_N)$. It can be shown
  \cite[Thm.~5.8]{Benkart1994} that the image of the walled Brauer
  algebra exhausts the centralizer, $\phi\bigl(\wBr(N)\bigr)\cong
  Z(\gl_N)$ (this is true for any $N$) and that $\phi$ is an
  isomorphism for $N\ge L$, so that $Z(\gl_N)\cong\wBr(N)$ in that
  case. For $N<L$ on the other hand, the commutant $Z(\gl_N)$ exhibits
  more relations as compared to $\wBr(N)$. Indeed, in this parameter
  range every simple basis element $e_{a_1}\otimes e^{a_2}\otimes
  \cdots e_{a_{L-1}}\otimes e^{a_{L}}$ of
  $(\cV\otimes\dual{\cV})^{\otimes L/2}$ has at least two
  subscripts which are identical, and attempts to antisymmetrize them
  will result in zero. (Here $e_a$ and $e^a$ stand, respectively, for
  a basis of $\cV$ and its dual.) This means that the map $\phi$
  representing the walled Brauer algebra has a non-trivial kernel, so
  that the representation is not faithful.

  Let us now study the implications of the previous statements. For
  general values of $L$ and $N$, the relevant decomposition of the
  Hilbert space reads
\begin{align}
  \label{eq:VVstar_gl_z}
  (\cV\otimes\dual{\cV})^{\otimes L/2}\big|_{\gl_N\otimes Z(\gl_N)}
  \ \cong\ \bigoplus_{K=0,2}^L\,\sideset{}{'}\bigoplus_{\mu,\nu\,\vdash K/2}
  V\bigl([\mu,\nu]_N\bigr)\otimes\mathcal{Z}_L^{\mu,\nu}\, ,
\end{align}
  where $\mathcal{Z}_L^{\mu,\nu}$ are certain representations of
  $Z(\gl_N)$. According to Eq.~\eqref{eq:glweight} it is required to
  restrict the summation to pairs of partitions satisfying
  $\ell(\mu)+\ell(\nu)\le N$ in order to ensure the existence of a
  bona fide weight $[\mu,\nu]_N$. This is indicated by the prime. Due
  to the existence of the homomorphism $\phi:\wBr(N)\to Z(\gl_N)$, the
  spaces $\mathcal{Z}_L^{\mu,\nu}$ can also be regarded as
  representations of $\wBr(N)$. For $N\geq L$ one has
  $Z(\gl_N)\cong\wBr(N)$ and
  $\mathcal{Z}_L^{\mu,\nu}\cong\mathcal{W}_L^{\mu,\nu}$. On the other
  hand, it may occur that the dimension $z_{\mu,\nu}$ of
  $\dim\mathcal{Z}_L^{\mu,\nu}$ is strictly smaller than the dimension
  $d_L^{\mu,\nu}$ of the representation $\mathcal{W}_L^{\mu,\nu}$.  A
  precise condition for this to happen has been given in
  \cite[Thm.~2.14]{Benkart1994}. It states that $z_{\mu,\nu}\le
  d_L^{\mu,\nu}$ for all $(\mu,\nu)\vdash (K/2,K/2)$, and
  $z_{\mu,\nu}= d_L^{\mu,\nu}$ if and only if
\begin{align}
  \label{eq:cond_N_munu}
  N\ \ge\ \ell(\mu)+\ell(\nu)+(L-K)/2\, .
\end{align}
  In these cases one encounters a mismatch between the spectrum of the
  spin model and the spectrum of the loop model since the latter is
  obtained by diagonalizing the Hamiltonian on the larger spaces
  $\mathcal{W}_L^{\mu,\nu}$. In practice, certain energy eigenvalues
  simply have to be discarded (we comment on this below).

  It is instructive to illustrate this result with a concrete
  example. Let us set $L=4$ and consider the decomposition of the
  Hilbert space for
\begin{equation}
\begin{split}
  N\geq4:\qquad
  (\cV\otimes\dual{\cV})^2
  &\ \cong\ 
  2 V\bigl([\,0^{N}\,]\bigr)
  \oplus
  4 V\bigl([1,0^{N-2},-1]\bigr)
  \\
  &\qquad\oplus  V\bigl([2,0^{N-2},-2]\bigr)
  \oplus
  V\bigl([2,0^{N-3},-1^2]\bigr)\\[2mm]
  &\qquad
  \oplus
  V\bigl([1^2,0^{N-3},-2]\bigr)
  \oplus
  V\bigl([1^2,0^{N-4},-1^2]\bigr)\ \ .
\end{split}
\end{equation}
  Here, the first term corresponds to $K=0$, the next to $K=2$ and the
  last four to $K=4$. For smaller values of $N$ one instead has
\begin{equation}
\begin{split}
  N=3:\qquad
  (\cV\otimes\dual{\cV})^2
  &\ \cong\ 
  2 V\bigl([\,0^{3}\,]\bigr)
  \oplus
  4 V\bigl([1,0,-1]\bigr)
  \\
  &\qquad\oplus  V\bigl([2,0,-2]\bigr)
  \oplus
  V\bigl([2,-1^2]\bigr)
  \oplus
  V\bigl([1^2,-2]\bigr)\ \ ,\\[2mm]
  \label{eq:VVstarN2}
  N=2:\qquad
  (\cV\otimes\dual{\cV})^2
  &\ \cong\ 
  2 V\bigl([\,0^{2}\,]\bigr)
  \oplus
  3 V\bigl([1,-1]\bigr)
  \oplus  V\bigl([2,-2]\bigr)\ \ .  
\end{split}
\end{equation}
  We see in particular that the multiplicity of $V\bigl([1,-1]\bigr)$
  is reduced when $N=2$.  This can be quickly checked by computing the
  dimension of each term on the r.h.s.~of Eq.~\eqref{eq:VVstarN2}. The
  space $V\bigl([1,-1]\bigr)$ can be identified with the $\SU(2)$
  representation of spin $1$ and $V\bigl([2,-2]\bigr)$ with that of
  spin $2$, so that the dimension is $2\times 1 + 3\times 3 + 1\times
  5=16$, coinciding indeed with the dimension of the Hilbert
  space. How the multiplicity is reduced in the general case if
  condition \eqref{eq:cond_N_munu} is not met has also been described
  in \cite{Benkart1994}, but the algorithm to compute it is quite
  complex and we do not describe it here.

  Let us finally briefly comment on a second source for a mismatch
  between the spectrum of the spin model and that of the loop
  model. In the decomposition \label{eq:VVstar_gl_z} of the spin
  Hilbert space we encountered a restriction to pairs of partitions
  satisfying $\ell(\mu)+\ell(\nu)\le N$. This constraint has no
  counterpart in the loop model. Of course, the resulting additional
  eigenvalues are under complete control and can easily be eliminated
  in the process of computing the spectrum.

  Altogether we now got a fairly complete picture of how the loop
  model can be used for studying the spin chain. As we argued above,
  the spectra are absolutely identical when $N\ge L$. More
  importantly, even for general values of $L$ and $N$ we expect a
  faithful representation of the spectrum in all sectors satisfying
  condition \eqref{eq:cond_N_munu}. We have checked these statements
  numerically.  In particular, we have verified for $N=2,3,4,5,6$ and
  $L=4,6$ that the lowest eigenvalues which will be relevant in
  Section~\ref{sc:Numerics} are present in the $\SU(N)$ spin chains.
  We remark that the missing eigenvalues of the loop model arise in a
  supersymmetric generalization of the spin chains at hand, with
  symmetry $\GL(N+M|M)$ and $M\ge 0$. This fact has been discussed for
  short-range spin chains in \cite{Candu:2009pj,Candu:2009ep}. The
  consequences for our long-range models will be addressed in a future
  publication \cite{Jochen:2014}.

\begin{figure}[h]
    \centering
    \begin{tikzpicture}
      \foreach \x in {0,2} { \draw[postaction={decorate,decoration={
            markings, mark=at position 0.45 with 
            {\arrow[scale=2]{>}}}}] (.7*\x,0) -- (.7*\x+.7*2,2); }
      \foreach \x in {3,5} { \draw[postaction={decorate,decoration={
            markings, mark=at position 0.45 with 
            {\arrow[scale=2]{>}}}}] (.7*\x,2) -- (.7*\x-.7*2,0); }
      \draw[postaction={decorate,decoration={ markings, mark=at
          position 0.26 with {\arrow[scale=2]{>}}}}] (.7*4,0) --
      (.7*0,2); \draw[postaction={decorate,decoration={ markings,
          mark=at position 0.26 with {\arrow[scale=2]{>}}}}] (.7*1,2)
      -- (.7*5,0);
      \foreach \x in {0,1,2,3,4,5} { \draw[fill] (.7*\x,0) circle
        (2pt); \draw[fill] (.7*\x,2) circle (2pt); }
      \draw (0,1) node[left] {$u^2\ =\quad$};
    \end{tikzpicture}
    \caption{\label{fig:u2}The diagram $u^2$ that translates two sites
      to the right for a system of size $L=6$. (Recall that periodic
      boundary conditions are imposed.)}
\end{figure}
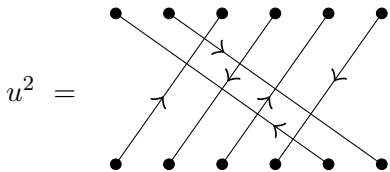

  Up to now we have only considered the on-site symmetry $\gl_N$ and
  its commutant $Z(\gl_N)$, the image of the walled Brauer algebra
  $\wBr(N)$. For the diagonalization problem it is, however, also
  useful to keep track of other conserved charges that commute with
  the Hamiltonian.  In particular, the dimension of the blocks of the
  Hamiltonian can be further reduced by exploiting its translational
  symmetry. For an equidistant arrangement of spins on the circle the
  Hamiltonian obviously commutes with the operator $u$ that implements
  the shift $\vec{S}_i\mapsto\vec{S}_{i+1}$. However, since the
  translation by a single site exchanges the roles of $\cV$ and
  $\dual{\cV}$ it is more appropriate in our context to work with the
  translation by two sites. Indeed, in contrast to $u$ the operator
  $u^2$ admits a natural interpretation as an element of the walled
  Brauer algebra $\wBr$. Its corresponding diagram is depicted in
  Figure~\ref{fig:u2}.

  The eigenvalues of $u^2$ are of the form $e^{4\pi i s/L}$, where the
  integer $s$ is defined modulo $L/2$ and coincides with the
  momentum. The reduced Hamiltonian $H_s\equiv\Pi_sH=H\Pi_s$ acting on
  the eigenspace of momentum $s$ is defined in terms of the projector
\begin{align}
  \Pi_s= \frac{2}{L}\sum_{t=0}^{L/2-1} e^{-4\pi i t s/ L }u^{2t}
  \quad\text{ which satisfies }\quad
  \Pi_s^2=\Pi_s\ \ .
\end{align}
  The eigenvalues of the Hamiltonian will then be labeled by the
  representations of the walled Brauer algebra and by their momentum.

  Before concluding this section, we give a brief historical note on
  the walled Brauer algebra.  This algebra was introduced in the
  mathematical literature precisely to study the problem of
  decomposing the space formed out of mixed products of fundamental
  and dual representations of the general linear group. One says that
  the general linear group and the walled Brauer algebra are in
  Schur-Weyl duality on these mixed tensor products, since their
  actions mutually centralize each other. This generalizes the well
  known Schur-Weyl duality between the symmetric group and the general
  linear group which applies when both are acting on the tensor
  product of fundamental representations only.  The latter corresponds
  to the setup of the uniform chain discussed in
  Section~\ref{sc:Uniform}. The walled Brauer algebra is a subalgebra
  of the Brauer algebra \cite{Brauer1937}.  The latter is in
  Schur-Weyl duality with the orthogonal group for the action in the
  tensor product of fundamental representations. The name ``walled''
  comes from the fact that one usually considers diagrams where the
  first $L/2$ lines on the left are directed upwards and the remaining
  $L/2$ lines on the right are directed downwards. With this
  convention up and down lines are separated by a domain
  wall which can only be crossed by horizontal arcs.
  Our setting is simply obtained by rearranging the order of lines.

\subsection{\label{sc:Numerics}Numerical study}

  In this section we will discuss the low energy properties of the
  spin chains and loop models.  Before entering the discussion for
  general $N$, it is useful to recall what happens for $N=2$. In that
  case, the fundamental and the anti-fundamental representation are
  equivalent and the symmetric rank-three tensor $d$ vanishes
  identically. This means that one is dealing with a uniform chain and
  our spin chain Hamiltonian is related to that of the Haldane-Shastry
  model,
\begin{align}
  \HHS\ =\ \sum_{i<j}  \frac{\vec{S}_i\cdot\vec{S}_j}
  {\sin^2\left(\smfrac{\pi}{L}(i-j)\right)}
  \qquad\text{ by }\qquad
  H(N=2)
  &\ =\ 4\,\HHS-\tfrac{2}{3}(L+1)\,\vec{S}^2
        +\tfrac{L}{3}(L^2+5)\ \ ,
\end{align}
  see our discussion in Section~\ref{sc:HamiltonianUniformED}. The
  Haldane-Shastry model is exactly solvable
  \cite{Haldane:1988PhRvL..60..635H,Shastry:1988PhRvL..60..639S}. In
  the continuum limit, the energies $E^{\text{HS}}_n$ of the
  low-lying states are given by the scaling dimensions $\Delta_n$ of
  the $\SU(2)$ WZW theory at level $k=1$:
  $(E^{\text{HS}}_n-E^{\text{HS}}_0)/(2L)=\Delta_n+O(1/L)$.  (The
  unusual power of $L$ comes from the dependence of the coupling on
  the length.) We remark that although the eigenstates of the
  Hamiltonian $H(N=2)$ are in correspondence with the fields in the
  WZW model, the universal part of the energy of those states is
  shifted due to the presence of the global Casimir operator
  $\vec{S}^2$.

  We now return to the case of general $N$ and use a numerical
  implementation of the loop model to determine the scaling properties
  of the energy gaps. If the low energy theory describing our model is
  a conformal field theory, one expects that
\begin{align}
  \label{eq:vsDelta}
  \frac{E_n-E_0}{2L}
  \ =\ v_s \Delta_n + O(1/L)\ \ ,
\end{align}
  where $\Delta_n=h+\bar{h}$ is the scaling dimension of the
  associated state. Here, we introduced the speed of sound $v_s$ to
  account for a possible numerical normalization factor which is
  independent of the energy level $n$ (it may depend on $N$
  though). Furthermore, as usual we identify the momentum $s$ of the
  state with the conformal spin: $s=h-\bar{h}$. Note that since the
  ground state energy of our model is zero, the gaps coincide with the
  energy of the excited states. In a CFT, the ground state energy
  scales with the length, with a prefactor proportional to the central
  charge of the theory. Unfortunately, we do not know how to shift our
  ground state energy in order to extract the central charge, and we
  will therefore focus on the spectrum of excited states only. We also
  recall that if instead our Hamiltonian is gapped, the energy
  difference between the first excited state and the ground state
  would behave as $(E_1-E_0)/(2L)\sim L$.

  In the following we will use the notation $E_{K,s,\ell}$ for the
  energy of the $\ell$-th excited state in the sector with $K$
  non-contractible strings and momentum $s$. The first excited state
  in the loop model occurs in the sector with $(K,s)=(2,0)$. In order
  to verify the absence of a gap for our Hamiltonian it is thus
  sufficient to study the scaling of the gap
  $E_{2,0,0}-E_{0,0,0}$. Specifically, we fit our data against the
  following function of $L$:
\begin{align}
  \label{eq:c0}
  \frac{E_{2,0,0}-E_{0,0,0}}{2L}
  \ =\ c_0\,L + c_1 + \frac{c_2}{L}\ \ .
\end{align}
  By definition of the gap one has $c_0\ge 0$. Table~\ref{tab:c0}
  summarizes the resulting values of $c_0$ determined in this way. We
  studied the values $N\in\{2,3,\ldots,10\}$, and in all these cases
  we found $c_0<10^{-2}$, which gives strong support to the hypothesis
  that the system is gapless and conformal for any $N\in\Integer_{\ge
    2}$. In particular, $c_0$ turns out to be exactly zero for $N=2$,
  since finite size corrections are practically absent in the
  Haldane-Shastry model. In what follows we will build on the
  conformal hypothesis to extract some conformal dimensions of the
  theory.

\begin{table}[h!c] 
  \centering 
  \begin{tabular}{|c|c|} 
  \hline 
  $N$ & $c_0$\\ 
  \hline 
  $2$ & $0$\\ 
  $3$ & $0.00448 \pm 0.00419$\\ 
  $4$ & $0.00669 \pm 0.00760$\\
  $5$ & $0.00460 \pm 0.00880$ \\ 
  $6$ & $0.00150 \pm 0.00958$\\ 
  $7$ & $0.00221 \pm 0.01263$\\ 
  $8$ & $0.00142 \pm 0.01486$\\ 
  $9$ & $0.00030 \pm 0.01695$ \\ 
  $10$ & $0.00050\pm 0.01980$\\
  \hline 
  \end{tabular} 
  \caption{\label{tab:c0}The coefficient $c_0$
  obtained from fitting Eq.~\eqref{eq:c0} for system sizes
  $L\in\{6,8,\dots,16\}$. The error is the standard error from the fit.
  }  
\end{table} 

  We first address the problem of determining the speed of sound $v_s$
  entering the scaling of the gaps \eqref{eq:vsDelta}. This is crucial
  in order to be able to extract the spectrum in the correct conformal
  units. We use the following argument. The gaps of our model (at
  least in the range $N\in\Integer_{\ge 2}$ we are interested
  in) are positive, and the ground state is identified in the CFT with
  the identity field, for which $h=\bar{h}=0$. The CFT state with
  $(h,\bar{h})=(2,0)$ is the holomorphic stress tensor which is always
  present in a CFT. If the spectrum is positive, then it {\em always}
  corresponds to the lowest state with conformal spin $s=2$. Moreover,
  it is an $\SU(N)$ singlet and should hence appear in the sector $K=0$
  of the loop model. Indeed, our numerics confirm that the lowest
  state with momentum $s=2$ occurs in the $K=0$ sector of the loop
  model, and we therefore expect that
\begin{align}
  \frac{E_{0,2,0}-E_{0,0,0}}{2L}
  \ =\ 2v_s + O(1/L)\ \ .
\end{align}
  Determining these gaps will enable us to determine $v_s$. The speed
  of sound measured as a function of $N$ is reported in
  Table~\ref{tab:vs}.
  We note that for $N=2,3,4$ the speed of sound is $N+2$ within the
  errors bars, while for larger $N$ it deviates from this value. At the
  same time the uncertainties also increase with $N$, showing that these
  deviations may be due to finite size corrections.\footnote{
  We note that normalizing energies as above, the speed of sound for
  the uniform spin chain of
  Eq.~\eqref{eq:modHaldaneShastryEquidistant} is precisely $N+2$, as
  known from the solution of the Haldane-Shastry model
  \cite{Haldane:1994cond.mat..1001H}.}  As a consistency
  check, we also determined $v_s$ by looking at the gaps
  $(E_{0,1,0}-E_{0,0,1})/(2L)$ between the first excited state with
  $(K,s)=(0,1)$ and the second excited state with $(K,s)=(0,0)$.  This
  energy difference is exactly $v_s$ if the state
  corresponding to $E_{0,1,0}$ is a descendant at level one of that
  corresponding to $E_{0,0,1}$. These states have lower energy than
  $E_{0,2,0}$  and they are less sensitive to finite size corrections.
  However, the speed of sound extracted from
  $(E_{0,1,0}-E_{0,0,1})/(2L)$ does not deviate significantly from the
  one presented in Table~\ref{tab:vs}. In the following we will
  therefore continue to use the values $v_s$ from the latter
  table. Our findings can also be viewed as a confirmation that the
  state $E_{0,1,0}$ indeed corresponds to a descendant of $E_{0,0,1}$.

\begin{table}[h!c]
  \centering
  \begin{tabular}{|c|c|c|c|}
    \hline
    \backslashbox{$L$}{$N$} & $2$ & $3$ & $4$\\
    \hline
    $6$ & $2$       & $2.60417$ & $3.2$     \\
    $8$ & $2.5$     & $3.27199$ & $4.03$    \\
    $10$ & $2.8$    & $3.65978$ & $4.50023$ \\
    $12$ & $3$      & $3.90832$ & $4.79233$ \\
    $14$ & $3.14286$& $4.07846$ & $4.98525$ \\
    $16$ & $3.25$   & $4.2005$  & $5.11822$ \\
    $16$ & $3.33333$& $4.29114$ & $5.21265$ \\
    \hline
    $\infty$ & $4$  & $5.0144\pm 0.01613$ & $5.96207\pm 0.03390$ \\
    \hline
  \end{tabular}\\[1cm]
  \begin{tabular}{|c|c|c|c|}
    \hline
    \backslashbox{$L$}{$N$} & $5$& $6$& $7$ \\
    \hline
    $6$ &  $3.79167$ & $4.38095$ & $4.96875$ \\
    $8$ &  $4.7811$  & $5.52829$ & $6.27304$ \\
    $10$ & $5.33125$ & $6.15699$ & $6.97948$ \\
    $12$ & $5.66463$ & $6.53046$ & $7.39236$ \\
    $14$ & $5.87838$ & $6.76405$ & $7.64524$ \\
    $16$ & $6.02061$ & $6.91475$ & $7.80397$ \\
    $16$ & $6.11744$ & $7.01335$ & $7.90404$ \\
    \hline
    $\infty$& $6.8801\pm 0.05210$ & $7.78294\pm 0.07036$ 
    & $8.67716\pm 0.08858$ \\
    \hline
  \end{tabular}\\[1cm]
  \begin{tabular}{|c|c|c|c|}
    \hline
    \backslashbox{$L$}{$N$} & $8$ & $9$& $10$\\
    \hline
    $6$  & $5.55556$ & $6.14167$ & $6.72727$  \\
    $8$  & $7.0162$  & $7.75823$ & $8.49945$  \\
    $10$ & $7.79984$ & $8.61872$ & $9.43654$  \\
    $12$ & $8.25169$ & $9.10926$ & $9.96557$  \\
    $14$ & $8.52353$ & $9.39984$ & $10.2747$  \\
    $16$ & $8.69006$ & $9.57403$ & $10.4565$  \\
    $18$ & $8.79142$ & $9.67659$ & $10.5602$ \\
    \hline
    $\infty$ & $9.56609\pm 0.10672$ & 
    $10.4516\pm 0.12479$ & $11.3347\pm 0.14280$ \\
    \hline
  \end{tabular}
  \caption{\label{tab:vs}The rescaled gap $(E_{0,2,0}-E_{0,0,0})/(4L)$
    as a function of $N$ and $L$. The row at $L=\infty$ is the
    extrapolated value of the speed of sound $v_s$ obtained by fitting
    the data with a polynomial in $1/L$ of degree $2$. The error is
    the standard error from the fit.
  }
\end{table}

\begin{figure}[h!]
  \centering
\includegraphics[scale=1]{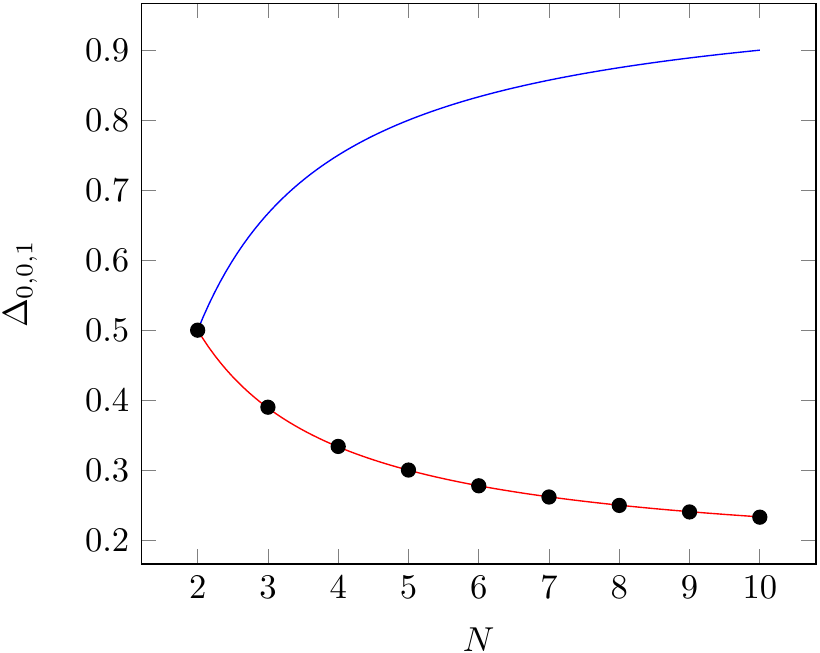}
\caption{The black points are the scaling dimension
  $\Delta_{0,0,1}$ of the first excited state in the sector $K=0$ as a
  function of $N$.  The lower solid curve is a fit of the data by
  $(N+4)/(6 N)$, while the upper dashed curve depicts the function
  $(N-1)/N$, the scaling dimension of the fundamental field in the
  $\SU(N)_1$ WZW model.  }
  \label{fig:delta_vs_N}
\end{figure}

  Having determined the speed of sound we can now estimate the lowest
  conformal dimensions of the CFT describing our model. As remarked at
  the beginning of this section for the $N=2$ case, we expect that in
  general the universal part of our Hamiltonian will be the sum of a
  CFT Hamiltonian plus non-local terms which shift the CFT conformal
  dimensions extracted from finite size scaling.  Due to the symmetry
  of our spin chain Hamiltonians these non-local terms have to
  correspond to global $\SU(N)$ Casimir operators. The general form of
  the resulting theory for $L\to\infty$ is then
\begin{align}
  H\ =\ H_{\text{CFT}}+\lambda_1+\lambda_2\,\vec{S}^2
        +\lambda_3\,\vec{S}^3+\cdots+\lambda_N\,\vec{S}^N\ \ ,
\end{align}
  see also Eq.~\eqref{eq:HChemicalPotential}. The aforementioned
  shifts will not be present in energy differences of states carrying
  the same representation. In particular, this is the case for the
  gaps in the singlet sector, where all Casimir invariants act
  trivially. In the following we shall hence focus on the subspace of
  singlets, corresponding to the sector $K=0$ of the loop model
  (absence of non-contractible lines), and determine the scaling of
  the first excited state. The latter has momentum $s=0$, and we
  denote its energy by $E_{0,0,1}$. The results for the extracted
  dimensions are presented in Table~\ref{tab:delta001}. We find that
  the measured values of $\Delta_{0,0,1}$ are well described by the
  function $(N+4)/(6 N)$. In Figure~\ref{fig:delta_vs_N} we plot
  $\Delta_{0,0,1}$ (black points) against this function (solid curve)
  and the smallest positive scaling dimension of a $\SU(N)_1$ WZW
  singlet field, namely $(N-1)/N$ (dashed curve). We see that the
  dimensions extracted are not consistent with those predicted by the
  $\SU(N)_1$ WZW model for $N>2$.
  We cannot exclude the possibility that the scaling dimensions are not
  exactly described by $(N+4)/(6N)$ since, as remarked above, the
  results for larger $N$ ($>4$) are less reliable due to finite size
  effects.
  Irrespective of their exact (but unknown) values, we note a
  clear tendency in our data: The measured dimensions decrease with
  $N$ while those of the WZW model increase.  
  This finding provides a strong indication that the CFT describing
  the alternating spin chains for $N>2$ is different from the
  $\SU(N)_1$ WZW model. The identification of this theory can be tackled using
  the methods presented in this section. Our approach even allows us
  to study the more general setup of loop models with an arbitrary
  value of the fugacity $\delta$. We relegate a detailed study of the
  resulting CFTs to another publication \cite{Jochen:2014}, where we
  will present a more general point of view based on supersymmetric
  spin chains.

\begin{table}[h!c]
  \centering
  \begin{tabular}{|c|c|}
    \hline
    $N$ & $\Delta_{0,0,1}$\\
    \hline
    $2$ & $0.5$  \\
    $3$ & $0.38956 \pm 0.00201$ \\
    $4$ & $0.33345\pm 0.00296$ \\
    $5$ & $0.29971\pm 0.00349$ \\
    $6$ & $0.27723\pm 0.00381$ \\
    $7$ & $0.26121\pm 0.00402$ \\
    $8$ & $0.24921\pm 0.00417$ \\
    $9$ & $0.23990\pm 0.00428$  \\
    $10$ & $0.23246\pm 0.00436$ \\
    \hline
  \end{tabular}
  \caption{\label{tab:delta001}The scaling dimension $\Delta_{0,0,1}$ of the 
    first excited state in the sector $K=0$ as a function of $N$.
    The values are extracted from fitting data for systems with 
    $L\in \{6,8,\dots,18\}$
    with a polynomial in $1/L$ of degree two, and using the
    numerically determined speed of sound.
  }
\end{table}

\section{\label{sc:Conclusions}Conclusions and Outlook}

  In this article we have constructed several families of long-range
  $\SU(N)$ spin models in 1D and 2D. They all arise as parent
  Hamiltonians for infinite matrix product states based on the
  $\SU(N)$ WZW model at level $k=1$. The whole construction is based
  on a given groundstate, which is known exactly and can be expressed
  in terms of chiral correlation functions of WZW primary fields. At
  level $k=1$ the latter can be evaluated explicitly using a free
  field realization. For simplicity we restricted our attention to
  spin models involving the fundamental and the anti-fundamental
  representation of $\SU(N)$ but the generalization to other
  representations should be straightforward.

  The models we constructed give rise to a 2D conformal field theory
  if the spins are placed equidistantly on a circle. If only the
  fundamental representation is involved, the Hamiltonian essentially
  reduces to the $\SU(N)$ Haldane-Shastry model up to the addition of
  non-local chemical potentials corresponding to global $\SU(N)$
  Casimir operators. In this case, the model admits an exact analytic
  solution and it flows to the $\SU(N)_1$ WZW model in the
  thermodynamic limit. The case of an alternating spin chain turns out
  to be more complicated. Our numerical analysis provides strong
  evidence that this long-range spin chain is critical as
  well. However, our results on the conformal spectrum rule out that
  the critical point is described by a $\SU(N)_1$ WZW model. Most of
  our analysis is based on a reformulation of the original problem in
  terms of loop models. These are not only providing the
  computationally most efficient representation of the Hamiltonian
  (both in 1D and 2D) but they are also interesting in their own
  right.

  With regard to the physical interpretation of our Hamiltonians it
  will be crucial to achieve a better understanding of different types
  of 2D setups and to relate them to the physics of fractional quantum
  Hall states and chiral spin liquids. For the case of the $\U(1)_k$
  WZW model a connection to Laughlin states could be established in
  \cite{Nielsen:2012zb,Tu:2014NJPh...16c3025T}. A similar analysis for
  $\SU(3)$ should result in a connection with the Halperin
  \cite{Halperin:1983} or variants of the non-abelian spin singlet
  (NASS) states
  \cite{Ardonne:1999PhRvL..82.5096A,Ardonne:2001NuPhB.607..549A}.

  An interesting technical problem regards the determination of
  spin-spin correlation functions for the models we have
  constructed. These could be used to substantiate any claim on the
  gapless or gapped nature of the resulting phases. In the case of
  $\SU(2)$ it was possible to derive a recursion relation for
  two-point functions which could then be solved systematically, both
  for the finite and the infinite equidistant chain
  \cite{Nielsen:2011py}. Similar recursion relations can be derived
  for $\SU(N)$. However, due to the existence of the non-trivial
  tensor $d_{abc}$ they now only relate two-point functions to
  three-point functions instead of giving an equation for the
  two-point function itself. As a consequence, the recursion relations
  can only be used to verify existing proposals but not to provide a
  solution from first principles. In view of existing conjectures
  about the dynamical spin-spin correlators in the $\SU(N)$
  Haldane-Shastry model
  \cite{Haldane:1994cond.mat..1001H,Yamamoto:2000JPSJ...69..900Y} the
  study of these recursion relations might nevertheless be an avenue
  worth pursuing.

  In our opinion, the most pressing open question concerns the nature
  of the critical theory arising from the alternating $\SU(N)$ spin
  chain on the circle discussed in
  Section~\ref{sc:HamiltonianAlternatingED}.  In the context of our
  reformulation in terms of loop models it is natural to revisit this
  question from a more general perspective. First of all, it is
  natural to regard the symmetry group $\SU(N)$ as a special instance
  of the family $\SU(N+M|M)$ of special unitary supergroups. This
  alternative point of view has the advantage that the spectrum of the
  loop model and that of the spin chain match precisely for
  sufficiently large values of $M$. Moreover, in the loop formulation
  the number $N$ can be regarded as a continuous parameter and it will
  be interesting to explore the different regimes where critical
  behavior can be expected. For instance, thanks to a mapping onto the
  $N^2$ states Potts model, it is known that standard loop models
  without crossings and with nearest neighbor interactions cease to be
  critical for fugacities $N>2$. Our numerical results in
  Section~\ref{sc:Loops} indicate that this bound is not relevant for
  our types of long-range crossing loop models. We plan to return to
  these issues in a forthcoming publication \cite{Jochen:2014}.

  Taking into account the results of this paper, infinite matrix
  product states based on WZW models have now been constructed for the
  symmetry groups $\U(1)$, $\SU(N)$ and $\SO(N)$
  \cite{Nielsen:2011py,Tu:2014NJPh...16c3025T,Tu:2013PhRvB..87d1103T}. The
  only remaining groups of classical type are the symplectic groups
  $\SP(2N)$. This case is currently under investigation and we hope to
  report on it in the near future.

\subsubsection*{\centerline{$\ast\ast\ast$ Note added in proof $\ast\ast\ast$}}

  During the preparation of this manuscript, we learned that related
  results have been obtained by Hong-Hao Tu, Anne Nielsen and German
  Sierra \cite{TNS:2014}.

\subsubsection*{Acknowledgements}

  We would like to thank Eddy Ardonne, Jochen Peschutter and Hong-Hao
  Tu for useful discussions and Hong-Hao Tu, Anne Nielsen and German
  Sierra for agreeing to a joint submission to the arXiv. The
  authors of this article are both funded by the German Research
  Council (DFG) through Martin Zirnbauer's Leibniz Prize, DFG grant no.\
  ZI 513/2-1. Additional support is received from the DFG through the
  SFB$|$TR\,12 ``Symmetries and Universality in Mesoscopic Systems''
  and the Center of Excellence ``Quantum Matter and Materials''.

\appendix
\section{\label{ap:SUN}Some basic facts on $\mathbf{SU(N)}$}

  Since many of the algebraic expressions in the main text involve the
  invariant tensors of $\SU(N)$ we find it useful to summarize a few of
  the most important formulas for them. In what follows, the symbol
  $T^a$ refers to the spin matrices in the fundamental representation
  $\cV$ in an arbitrary basis. The first object of interest is the
  metric which is defined by
\begin{align}
  \label{eq:SMetric}
  \kappa^{ab}\ =\ \tr_\cV(T^aT^b)\ \ .
\end{align}
  Throughout the text, the metric $\kappa^{ab}$ and its inverse
  $\kappa_{ab}$ are used to raise and lower indices. The structure
  constants ${f^{ab}}_c$ and ${d^{ab}}_c$ may then be introduced via
  the identity
\begin{align}
  \label{eq:TProduct}
  T^aT^b
  \ =\ \tfrac{1}{2}\bigl[T^a,T^b\bigr]
       +\tfrac{1}{2}\bigl\{T^a,T^b\bigr\}
  \ =\ \tfrac{i}{2}{f^{ab}}_c\,T^c
       +\tfrac{1}{2}{d^{ab}}_c\,T^c
       +\tfrac{1}{N}\kappa^{ab}\,\bI\ \ ,
\end{align}
  where the first term corresponds to the antisymmetric part and the
  remaining ones to the symmetric part. An alternative way of
  introducing these tensors is
\begin{align}
  f^{abc}
  \ =\ -i\tr_\cV\bigl([T^a,T^b]T^c\bigr)
  \qquad\text{ and }\qquad
  d^{abc}
  \ =\ \tr_\cV\bigl(\{T^a,T^b\}T^c\bigr)\ \ .
\end{align}
  By construction, these tensors are completely (anti)-symmetric,
  respectively. Both of them are traceless, i.e.\
  $\kappa_{ab}f^{abc}=\kappa_{ab}d^{abc}=0$. The tensor $d^{abc}$
  vanishes for $\SU(2)$ but it is non-trivial for all integers
  $N\geq3$. It remains to summarize a few identities involving two or
  three of these tensors,
\begin{align}
  \label{eq:TensorIdentities}
  {f^{ac}}_d\,{f^{bd}}_c
  &\ =\ -2N\kappa^{ab}\quad,&
  {f^a}_{ec}\,{f^e}_{bd}\,{f^{cd}}_g
  &\ =\ -N{f^a}_{bg}\quad,\\[2mm]
  {d^{ac}}_d\,{d^{bd}}_c
  &\ =\ \tfrac{2(N^2-4)}{N}\,\kappa^{ab}\quad,&
  {f^a}_{ec}\,{f^e}_{bd}\,{d^{cd}}_g
  &\ =\ -N{d^a}_{bg}\ \ .
\end{align}
  More relations of a similar type can be found in
  \cite[Page~92]{Bouwknegt:1993wg} and references therein.

  For the discussion of the alternating chain it is also necessary to
  have some information on the anti-fundamental representation
  $\dual{\cV}$. The corresponding representation matrices $\dual{T}^a$
  are related to those in the fundamental representation by
  transposition, $\dual{T}^a=-(T^a)^T$.  This definition together with
  \eqref{eq:TProduct} then immediately implies a product formula of
  the form
\begin{align}
  \label{eq:TDualProduct}
  \dual{T}^a\dual{T}^b
  \ =\ \tfrac{1}{2}\bigl[\dual{T}^a,\dual{T}^b\bigr]
       +\tfrac{1}{2}\bigl\{\dual{T}^a,\dual{T}^b\bigr\}
  \ =\ \tfrac{i}{2}{f^{ab}}_c\,\dual{T}^c
       -\tfrac{1}{2}{d^{ab}}_c\,\dual{T}^c
       +\tfrac{1}{N}\kappa^{ab}\,\bI\ \ .
\end{align}
  We note that there is an important sign difference as compared to
  the analogous expression \eqref{eq:TProduct} for the fundamental
  representation $\cV$. In the unified language of
  Section~\ref{sc:NullVectors}, the two relations \eqref{eq:TProduct}
  and \eqref{eq:TDualProduct} may be compactly expressed as
  \eqref{eq:SProduct}.

\bibliographystyle{../utphysTQ}
\bibliography{bibliography}

\end{document}